\documentclass[12pt]{article}

\usepackage{amssymb}
\usepackage{amscd}
\usepackage{graphicx}
\usepackage{amsfonts}
\usepackage{amsmath}
\usepackage{latexsym}
\usepackage[all]{xy}
\usepackage{fancyhdr}
\usepackage{color}

\usepackage{latexsym,graphics,color}
\def\w{\wedge}

\def\be{\begin{equation}}
\def\ee{\end{equation}}
\def\D{\mathcal D}

\def\A{\mathcal A}
\def\B{\mathcal B}
\def\K{\mathcal K}

\def\E{\mathcal E}
\def\C{\mathcal C}

\def\L{\mathcal L}
\def\bc{\mathbb C}
\def\br{\mathbb R}
\def\R{\mathcal R}
\def\J{\mathcal J}
\def\F{\mathcal F}
\def\dg{\dagger}
\def\L{\mathcal L}
\def\si{\sigma}
\def\fp{{\rm Fun}(P)}
\def\d{\partial}
\def\jp{\frac{1}{2}}
\def\ri{{\mathrm i}}
\def\bos{\boldsymbol}
\def\lj{  L\otimes {\rm Id}}
\def\ld{ {\rm Id}\otimes L}

\def\I{{\rm Id}}

\def\Ad{{\rm Ad}}
\def\ka{\kappa}
\def\ad{{\rm ad}}

\definecolor{lila}{rgb}{1,0.2,0.9}
\definecolor{brown}{rgb}{0.5,0.3,0.3}
\definecolor{turquoise}{rgb}{0.2,0.9,0.7}
\definecolor{Orange}{rgb}{0.93,0.44,0}           
\definecolor{GrayBlue}{rgb}{0.35,0.4,0.62}       
\definecolor{SeafoamGreen}{rgb}{0.54,0.71,0.50}  

\definecolor{darkorange}{cmyk}{.20,.50,.80,0}
\definecolor{lightorange}{cmyk}{.07,.37,.65,0}
\definecolor{darkpeagreen}{cmyk}{.50,.30,.50,0}
\definecolor{lightpeagreen}{cmyk}{.22,.20,.40,0}

\def\ri{{\mathrm{i}}}                   %
\def\bC{{\mathbb C}}                    %
\def\1{{\mbox{\boldmath $1$}}}          %
\def\tr{\mathrm{tr\,}}                  %
\def\lm{\lambda}                        %
\def\jp{\frac{1}{2}}                    %
\def\al{\alpha}                         %
\def\ga{\gamma}                         %

\definecolor{spec}{rgb}{0.0, 0.26, 0.15}
\def\bx{\boldsymbol{x}}

\def\tbx{\boldsymbol{\tilde x}}

\begin{document}

\begin{flushright}
{}~
  
\end{flushright}

\vspace{1cm}
\begin{center}
{\large \bf Brief lectures on duality, integrability and deformations\footnote{Based on the lectures given in Mathematical Sciences Institute, ANU, Canberra, Simons Center for Geometry and Physics, Stony Brook, Tohoku Forum for Creativity, Sendai and in the
Santiago de Compostela "Integrability, dualities and deformations" webinar.}}

\vspace{1cm}

{\small
{\bf Ctirad Klim\v{c}\'{\i}k}
\\
Aix Marseille Universit\'e, CNRS, Centrale Marseille\\ I2M, UMR 7373\\ 13453 Marseille, France}
\end{center}

\vspace{0.5 cm}

\centerline{\bf Abstract}
\vspace{0.5 cm}
\noindent   We provide a pedagogical introduction to some aspects of   integrability, dualities and deformations of physical systems in 0+1 and in 1+1 dimensions.  In particular,  we concentrate  on the T-duality of point particles and strings as well as on the Ruijsenaars duality of finite many-body integrable models, we review the concept of the  integrability and, in particular, of the Lax integrability and we analyze the basic examples of the Yang-Baxter deformations of  non-linear $\sigma$-models.  The central mathematical structure which we describe in detail is the  $\E$-model  which is the dynamical system exhibiting all those three phenomena simultaneously. The last part of the paper contains original results, in particular a formulation of   sufficient conditions for strong integrability of non-degenerate $\E$-models.
  
  \vspace{2pc}

\noindent Keywords: integrable systems, nonlinear  sigma models, 
T-duality, Ruijsenaars duality

  \section{Introduction}
  A recent progress in the theory of integrable
  nonlinear $\sigma$-models in two space-time dimensions  has brought to light the  relevance of the so-called $\E$-models for the integrability story. This relevance is intriguing, because the original motivation  \cite{KS96a,KS97,K15} for the introduction of the $\E$-models was rather to understand the dynamics of the T-dualizable $\sigma$-models and the integrability came only later as a welcome bonus \cite{K09}. 
  
  \medskip
  
  The role of this review is to describe in a succinct manner how the stories of the T-duality and of the integrable
  deformations meet together on the playground  of the $\E$-models. On the top of it, we explain what is the Ruijsenaars duality in the many-body integrable systems  \cite{siru} and we also formulate some original results concerning   sufficient conditions guaranteeing that a given $\E$-model is strongly Lax integrable.
  
  \medskip
  
  The plan of the paper is as follows: in Section 2, we briefly explain what we mean by "duality" and then we concentrate on the T-duality of point particles and strings. In the  case of strings, we start with the well-known exposition of the Abelian T-duality, then we provide a reformulation of the story in terms
  of the $\E$-model based on an Abelian Drinfeld double and, finally, we release the condition of the Abelianity to recover the full-fledged Poisson-Lie T-duality. We describe briefly also the degenerate version of the $\E$-model dynamics which gives rise to the so called dressing cosets
  T-duality. In Section 3, we focus on the case of $0+1$ dimension, namely we give a precise definition of the integrability of a many-body dynamical
  system and we explain in this context what is the so called Ruijsenaars duality. In Section 4, we introduce a special case of  the so called Lax integrability, which, unlike the general integrability, can be easily generalized from $0+1$ to $1+1$-dimensions. In this context, we distinguish a weak and a strong Lax integrability, in the latter case one must specify also the so called $r$-matrix as we shall illustrate on the example of the principal chiral model. Finally, we devote Section 5 to the description of the  Yang-Baxter deformations of the principal chiral model. We first describe a prototypical one-parameter deformation known under the name of Yang-Baxter $\sigma$-model \cite{K09} and then we deal in detail with a three-parametric case of the so called bi-YB-WZ $\sigma$-model introduced in Ref.\cite{DHKM,K19,K20}. In the latter case, we give a simplified proof
  of the strong Lax integrability than that given in Ref.\cite{K20} and, as an original result, we give  sufficient conditions for a general $\E$-model to be strongly Lax integrable.

  \section{T-duality} Duality is a very large term relevant in a vast and very diversified amount of examples. Therefore, we have first to make more precise the conceptual framework which will allow us to express  what we mean by the duality in the present article.
  
  \medskip
  
Consider  a set of  "structures" $S$ and  associate to a given structure $a\in S$ a set $M(a)$ of "points of views". Looking at the structure $a$ from a point of view $m\in M(a)$, we may or we may not recognize some pattern $b$ from the set of "patterns" $W$. If we do recognize it, we refer to the point of view $m$   as "meaningful" and we say that the structure $a$ exhibits
the pattern $b$.

\medskip

Given a structure $a\in S$, it may happen that there is no meaningful point of view to look at it (i.e. $a$ exhibits no pattern from the predefined set $W$), which is, course, an utterly uninteresting case. It also happens, that there is just one meaningful point of view and since there is then a one-to-one correspondence between    the pattern and the structure one is tempted to treat those two notions interchangeably. However, interestingly, 
  for some structures there may exist more than one meaningful point of view.   In particular, if there are
precisely two meaningful points of views $m_1$ and $m_2$ making the structure $a$ to exhibit two different  patterns $b_1$ and $b_2$, we say that    $b_1$ and $b_2$  are mutually dual to each other (we speak about the "self-duality" of $b_1$, if $m_1\neq m_2$ but $b_1=b_2$).
If there are more than two meaningful points of view for the  structure $a$, we speak about a "plurality".
 
\medskip
 
{An amusing example\footnote{The pictures come from the address http://immenseworlds.blogspot.com/2007/10/reverse-pictures-same-upside-down.html}
}: The set $S$ of the  "structures" is the set of   {\it drawings}. Given a drawing $a\in S$, we can  look at it from {\it  different angles} (points of view). The set of patterns $W$ is the set of   {\it human faces}.  The point of view is meaningful, if we recognize in the drawing such a human face.

 \bigskip
 
 \centerline{\large{\textcolor{spec}{\bf Example of duality}}}
 
\begin{center}

  \includegraphics[scale=0.8]{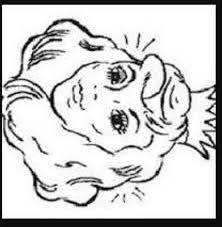}

A drawing from a non-meaningful point of view
   
\vskip2pc

\includegraphics[scale=0.8]{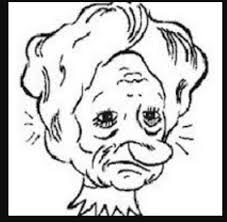}
\includegraphics[scale=0.8]{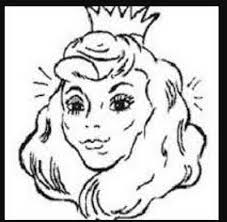}

Looking at the same drawing from two   meaningful points of view.  
 \end{center}
\vskip1pc
 
 \subsection{T-duality for point particles}
In this section, the set of "structures" will be the set
of  $2n$-dimensional {\it Hamiltonian dynamical systems}, this is to say the set of pairs
$(P,H)$, where $P$ is a $2n$-dimensional symplectic manifold and $H$ a distinguished function on $P$ called the Hamiltonian.

\medskip

Given the  dynamical system $(P,H)$, the "point of view" is every choice of local (Darboux) coordinates  $x^i,p_i$, $i=1,\dots, n$ on $P$ in which the symplectic form takes the canonical form
 \be \omega=dp_i\wedge dx^i.\ee
 The point of view is meaningful if the Hamiltonian $H$ expressed in the Darboux coordinates  $x^i,p_i$ takes the form
\be  H=\jp g^{ij}(x) \left(p_i+eA_i(x)\right) \left(p_j+eA_j(x)\right)+e\phi(x),\label{emg}\ee
where $g^{ij}(x)$ is a non-degenerate symmetric  tensor field. 
The corresponding "pattern" exhibited by the dynamical system $(P,H)$ is then the
 electro-magnetic-gravitational background   $(\phi(x),A_i(x),g_{ij}(x))$ on the Riemannian manifold   defined by the requirements $p_i=0$, $i=1,\dots, n$, where $g^{ij}(x)$ is the inverse metric tensor.

\medskip

\noindent {\bf Remark 2.1:}
{\small Recall for completeness that writing down the Hamiltonian equation of motions for the Hamiltonian \eqref{emg} and eliminating the momenta ${p_i}$ gives the  geodesic equation in the presence of the Lorentz force
\be \ddot x^i+\Gamma^i_{jk}(x)\dot x^j\dot x^k+eg^{ij}(x)(F_{jk}(x)\dot x^k+\d_j\phi(x))=0,\label{cg}\ee\be \Gamma^i_{jk}=\jp g^{im}\left(\d_k g_{mj}+\d_j g_{mk}-\d_m g_{jk}\right), \quad F_{jk}=\d_jA_k-\d_kA_j.\ee
The charged geodesics equation \eqref{cg} can be obtained  also  from a second order action  
\be {  S[x(t)]= \jp\int dt \ g_{ij}(x)\dot x^i\dot x^j -e\int dt \  (A_i(x)\dot x^i+\phi(x)).}\ee}
 
\medskip

The question we wish to ask is the following one: If the dynamical system $(P,H)$   admits Darboux coordinates yielding the meaningful form \eqref{emg} of its Hamiltonian $H$, can it exist another meaningful  point of view, i.e. 
another system of Darboux coordinates $\tilde x^i,\tilde p_i$ which would also yield
  the meaningful form \eqref{emg}  of the Hamiltonian but would exhibit a {\it different} background geometry $(\tilde\phi(\tilde x), \tilde A_i(\tilde x), \tilde g_{ij}(\tilde x))$? 
  
  \medskip
  
  To answer this question, we first   remark that if the Darboux coordinates 
  $\tilde x^i,\tilde p_i$ are related to the Darboux coordinates $x^i,p_i$ by a point-like
  canonical transformation 
  \be \tilde x^j=\tilde x^j(x),\quad \tilde p_j=\frac{\d x^k}{\d \tilde x^j}p_k,\label{201}\ee
  then the meaningful Hamiltonian \eqref{emg}
  becomes meaningful also in the  coordinates $\tilde x^i,\tilde p_i$ where it reads
 \be H =\jp \tilde g^{ij}(\tilde x) (\tilde p_i+e\tilde A_i(\tilde x)) (\tilde p_j+e\tilde A_j(\tilde x))+e\tilde\phi(\tilde x),\ee
\be \ \tilde g^{ij}(\tilde x)=\frac{\d \tilde x^i}{\d x^k}\frac{\d \tilde x^j}{\d x^l}g^{kl}(x), 
\quad \tilde A_i(\tilde x)=\frac{\d x^j}{\d \tilde x^i}A_j(x), \quad \tilde\phi(\tilde x)=\phi(x).\label{trans}\ee
However, the duality realized by the point-like transformation \eqref{201} is not an interesting one
 because the transformation formulas
\eqref{trans} give geometrically  the same electro-magnetic-gravitational background induced by  the diffeomorphism  $\tilde x^j=\tilde x^j(x)$.  

\medskip

To find a nontrivial duality, the original Darboux coordinates $x^i,p_i$ and the dual ones  $\tilde x^i,\tilde p_i$ must be therefore related by a canonical transformation which is not the point-like one
\eqref{201}. An example of such nontrivial transformation was given in \cite{K20} and we describe it here in detail.

\medskip

Consider a four-dimensional manifold $P=\br^2_*\times\br^2_*$, where $\br^2_*$ stands for two-dimensional plane without origin. Introducing two copies of the standard polar coordinates $r,\phi$ and $\rho,f$ on $\br^2_*$, we consider the following  four-parametric electro-gravitational background $T(\mu,\ga,m,c)$
 on $M$ 
\begin{subequations}\label{times}\begin{align}  
\varphi&=\jp\left(\ga^2+\frac{1}{r^2}\right)+\jp\left(c^2+\frac{1}{\rho^2}\right),\\ A&=0,\\ds^2 &=\frac{1}{1+\mu^2 r^{2}}dr^2  +   \frac{r^{2}}{1+\ga^2 r^{2}}  d\phi^2+\frac{1}{1+m^2 \rho^{2}}d\rho^2  +   \frac{\rho^{2}}{1+c^2 \rho^{2}}  df^2.\label{ta}\end{align}\end{subequations}

\smallskip
 
The dynamics of the charged point particle of the positive charge $e$ in the background $T(\mu,\ga,m,c)$ is then governed by the Hamiltonian
  \be H =  \jp(1+\mu^2 r^2)p_r^2+\frac{1+\ga^2 r^2}{2r^2} (p_\phi^2+e) +\jp(1+m^2 \rho^2)p_\rho^2+\frac{1+c^2\rho^2}{2\rho^2} (p_f^2+e) \label{H2t}\ee
  and by the symplectic form
  \be \omega=p_r\w dr+p_\phi\w d\phi+p_\rho\w d\rho+p_f\w df.\label{omt}\ee
  
 We now express the  coordinates $p_r,r>0,p_\rho,\rho>0, p_\phi,\phi,p_f,f$ on the phase space $P$  in terms of new coordinates $P_R,R>0,P_\R,\R>0, P_\Phi,\Phi,P_F,F$
 as follows
  \begin{subequations} \label{ct}\begin{align} r&=\R\frac{\sqrt{\R^2P^2_\R +P_\Phi^2+e}}{\sqrt{\R^2P^2_\R +P_F^2+e}},\quad p_r=P_\R\frac{\sqrt{\R^2P^2_\R +P_F^2+e}}{\sqrt{\R^2P^2_\R +P_\Phi^2+e}},\quad p_\phi=P_\Phi,   \\
 \rho&=R\frac{\sqrt{R^2P^2_R +P_F^2+e}}{\sqrt{R^2P^2_R +P_\Phi^2+e}}, \quad p_\rho=P_R\frac{\sqrt{R^2P^2_R +P_\Phi^2+e}}{\sqrt{R^2P^2_R +P_F^2+e}}, \quad 
p_f=P_F,\\
    f&=F+\frac{P_F}{\sqrt{P_F^2+e}}\arctan{\left(\frac{RP_R}{\sqrt{P_F^2+e}}\right)}-\frac{P_F}{\sqrt{P_F^2+e}}\arctan{\left(\frac{\R P_\R }{\sqrt{P_F^2+e}}\right)},\\
 \phi&=\Phi-\frac{P_\Phi}{\sqrt{P_\Phi^2+e}}\arctan{\left(\frac{R P_R}{\sqrt{P_\Phi^2+e}}\right)}+\frac{P_\Phi}{\sqrt{P_\Phi^2+e}}\arctan{\left(\frac{\R P_\R}{\sqrt{P_\Phi^2+e}}\right)}.\end{align}\end{subequations}
 The transformation \eqref{ct} is the diffeomorphism
 of the phase space $P$ with the inverse diffeomorphism given by
 {\begin{subequations} \label{cti}\begin{align} R&=\rho\frac{\sqrt{\rho^2p^2_\rho +p_\phi^2+e}}{\sqrt{\rho^2p^2_\rho +p_f^2+e}},\quad P_R=p_\rho\frac{\sqrt{\rho^2p^2_\rho +p_f^2+e}}{\sqrt{\rho^2p^2_\rho +p_\phi^2+e}},\quad P_\Phi=p_\phi,   \\
 \R&=r\frac{\sqrt{r^2p^2_r +p_f^2+e}}{\sqrt{r^2p^2_r +p_\phi^2+e}}, \quad P_{\R}=p_r\frac{\sqrt{r^2p^2_r +p_\phi^2+e}}{\sqrt{r^2p^2_r +p_f^2+e}}, \quad 
P_F=p_f,\\
    F&=f-\frac{p_f}{\sqrt{p_f^2+e}}\arctan{\left(\frac{\rho p_\rho}{\sqrt{p_f^2+e}}\right)}+\frac{p_f}{\sqrt{p_f^2+e}}\arctan{\left(\frac{r p_r}{\sqrt{p_f^2+e}}\right)},\\
 \Phi&=\phi+\frac{p_\phi}{\sqrt{p_\phi^2+e}}\arctan{\left(\frac{\rho p_\rho}{\sqrt{p_\phi^2+e}}\right)}-\frac{p_\phi}{\sqrt{p_\phi^2+e}}\arctan{\left(\frac{r p_r}{\sqrt{p_\phi^2+e}}\right)}.\end{align}\end{subequations}}

 Moreover,  the transformation \eqref{ct} is the symplectic diffeomorphism (or symplectomorphism) of the phase space $P$ because it preserves the symplectic form $\omega$. Indeed, inserting the formulas \eqref{ct} into \eqref{omt} gives
  \be \omega=dP_R\w dR+dP_\Phi\w d\Phi+dP_\R\w d\R+dP_F\w dF.\ee
  
It remains to show that the canonical transformation
\eqref{ct} can be interpreted as the T-duality symplectomorphism. For that, we express the Hamiltonian \eqref{H2t} in terms of the new Darboux coordinates $P_R,R>0,P_\R,\R>0, P_\Phi,\Phi,P_F,F$.
The result is  
  \be H = \jp(1+m^2 R^2)P_R^2+\frac{1+\ga^2R^2}{2R^2}(P_\Phi^2+e)+\jp(1+\mu^2 \R^2)P_\R^2+ \frac{1+c^2\R^2}{2\R^2}(P_F^2+e). \label{H2td}\ee
  
The  comparison of the formula \eqref{H2td} with Eq.\eqref{H2t} shows that the role of the parameters
$m$ and $\mu$ got exchanged while the parameters
$c$ and $\ga$ remained in their places. Said in other words, the Hamiltonian \eqref{H2td} describes
the dynamics of the charged point particle in the
dual background $T(m,\ga,\mu, c)$
\begin{subequations}\label{dtimes}\begin{align} \tilde\varphi &=\jp\left(\ga^2+\frac{1}{R^2}\right)+\jp\left(c^2+\frac{1}{\R^2}\right),\\\tilde A&=0,\\ \tilde ds^2 &=\frac{1}{1+m^2 R^{2}}dR^2  +   \frac{R^{2}}{1+\ga^2 R^{2}}  d\phi^2+\frac{1}{1+\mu^2 \R^{2}}d\R^2  +   \frac{\R^{2}}{1+c^2 \R^{2}}  dF^2
.\end{align}\end{subequations}

To conclude the argument that this point-particle T-duality  indeed does something non-trivial, it is sufficient to show that the
flipping of the parameters $\mu$ and $m$ may indeed alter
the Riemannian geometry of the dual background $T(m,\ga,\mu, c)$ with respect to that of the original one
$T(\mu,\ga,m, c)$. For that, consider for example the background $T(0,0,1,1)$ with the metric  
\be  ds^2 = dr^2  +   r^2  d\phi^2+  \frac{d\rho^2+\rho^{2}df^2}{1+\rho^{2}}.\label{or}\ee
We see that this is the Riemannian geometry of the direct product of the Euclidean plane with the Euclidean black hole \cite{Wi}.
 The metric corresponding to the dual background
 $T(1,0,0,1)$ is
 \be \tilde ds^2 =\frac{1}{1+ r^{2}}dr^2  +  r^2 d\phi^2+ d\rho^2  +   \frac{\rho^{2}}{1+  \rho^{2}}  df^2.\label{du}\ee
 We work out easily that 
 the respective Ricci scalars of the metrics \eqref{or} and \eqref{du} read
 \be Ric=\frac{4}{1+\rho^2},\quad \widetilde {Ric}=-2+\frac{6}{(1+\rho^2)^2}.\ee
 We thus observe  that the Riemannian geometries \eqref{or} and \eqref{du} are inequivalent, because
 $Ric$ is strictly positive while $\widetilde {Ric}$ acquires also negative values.

 \subsection{  T-duality for strings}
 The appropriate description of T-duality in string theory is achieved by choosing as the
  set of "structures"   the so called  loop dynamical systems, i.e. the triples $(LD,\omega,H)$,  where   $LD$ is the  space of loops (every point of the manifold $LD$  is a map from the unit circle  $S^1$ to some finite-dimensional     manifold  $D$),  $\omega$ a symplectic form on  $LD$
   and  $H$ is a distinguished function on  $LD$ called
   the  Hamiltonian.
 
 \medskip
 
 The "point of view" is a local system of     loop Darboux coordinates  $x^i(\sigma),p_i(\sigma)$ on   $LD$ in which the symplectic form takes the form
\be   \omega=\oint \delta p_i(\sigma)\wedge  \delta x^i(\sigma).\label{lsf}\ee
Note that the symbol $\oint$ stands for the integration over the loop parameter   $\sigma$ and, as usual,
$\delta$ denotes the de Rham exterior derivative in the infinite-dimensional context.

\medskip

   The point of view is meaningful if the Hamiltonian $H$  expressed in the loop Darboux coordinates  $x^i(\sigma),p_i(\sigma)$  takes the form
\be  H=\jp \oint g^{ij}(x)\left(p_i-b_{ik}(x )x'^k \right)\left(p_j-b_{jl}(x )x'^l \right) + \jp \oint g_{ij}(x)x'^ix'^j.\label{sh}\ee
Finally, the "pattern" exhibited by the dynamical system  $(LD,\omega,H)$  is  the  
 Kalb-Ramond-gravitational background  $(b_{ij}(x),g_{ij}(x))$ on the Riemannian manifold  $T$ spanned by the coordinates  $x^i$, $i=1,\dots, n$. 
 
 \medskip

  Eliminating the  coordinates ${   p_i(\sigma)}$ from the  Hamiltonian equations of motion  of the system \eqref{lsf}, \eqref{sh}
  gives the      geodesic surface  equation of the string exposed simultaneously to  the  gravitational and to the Kalb-Ramond Lorentz force:
$$\d_+\d_-x^i+\Gamma^i_{jk}(x)\d_+ x^j\d_-x^k+ \jp g^{il}(x)h_{ljk}(x)\d_+ x^j\d_-x^k=0.$$
Here ${\d_\pm=\d_\tau\pm\d_\sigma}$ and the Kalb-Ramond field strength ${  h_{ljk}(x)}$ is  the exterior derivative\footnote{The classical string equations make perfect sense also in the case where the three form ${  h_{ljk}(x)}$ is closed but is not exact, but we stick for the moment to discussion of the cases where the Kalb-Ramond potential $b_{ik}(x )$ exists globally.} of the $2$-form field ${ b_{ik}(x)}$:
\be h_{ljk}(x)=  \d_lb_{jk}(x)+\d_jb_{kl}(x)+\d_kb_{lj}(x).\label{260}\ee
The second-order action leading to the geodesic surface equation reads
 \be{  S[x(\tau,\sigma)]= \!\int  d\tau \oint \left(g_{ij}(x)+b_{ij}(x)\right)\d_+x^i\d_- x^j \equiv \!\int  d\tau \oint \ \! e_{ij}(x)\d_+x^i\d_- x^j .}\label{248}\ee

The crucial fact underlying the string T-duality story is as follows: There are examples of the loop dynamical systems
$(LD,\omega, H)$ admitting more than one meaningful points of view. This means that there are at least two sets
of the local Darboux coordinates, say
  $p_i(\sigma),x^i(\sigma)$ and   $\tilde p^i(\sigma),\tilde x_i(\sigma)$, such that the Hamiltonian of this system exhibits simultaneously two geometric patterns
$$  H=\jp \int d\sigma g^{ij}(x)\left(p_i-b_{ik}(x )x'^k \right)\left(p_j-b_{jl}(x )x'^l \right) + \jp \int d\sigma g_{ij}(x)x'^ix'^j=$$
\be =\jp \int d\sigma \tilde g_{ij}(\tilde x)\left(\tilde p^i-\tilde b^{ik}(\tilde x )\tilde x'_k \right)\left(\tilde p^j-\tilde b^{jl}(\tilde x )\tilde x'_l \right) + \jp \int d\sigma \tilde g^{ij}(\tilde x)\tilde x'_i\tilde x'_j\ee
where the geometric data  ${ b_{ik}(x),g_{ik}(x)}$  need not be related by a general coordinate transformation to the geometric data  ${ \tilde b^{ik}(\tilde x),\tilde g^{ik}(\tilde x)}$.

\medskip

\noindent {\bf Remark 2.2:} {\small The flipping of the dual indices upside down is against the usual conventions but this notation is often very practical in the T-duality business.}

\medskip

\noindent {\bf Remark 2.3:} {\small An equivalent description of the T-duality phenomenon is as follows: consider two manifolds $T$ and $\tilde T$, each one equipped with the metric and with the Kalb-Ramond field. They are said to be T-dual to each other if   the  associated string  dynamical systems are non-trivially equivalent. The non-trivial equivalence means the existence
of a symplectomorphism between the phase spaces of the strings moving in the geometry $T$ and in the geometry $\tilde T$ which is not the point canonical transformation. Moreover, this symplectomorphism must take the $\tilde T$-Hamiltonian into the  $T$-Hamiltonian.}
 
\subsection{Abelian T-duality}
The prototypical example of the string T-duality is the so called Abelian T-duality \cite{KY,SSe}. Topologically, the target manifold $T$ is the  direct product of $n$ circles   $T= S^1\times S^1\times \dots \times S^1$, the $i^{th}$ circle is parametrized by an angle $x^i\in[0,2\pi[$. The metric $  g_{ij}$ and the Kalb-Ramond field $b_{ij}$ on $T$ are taken to be constant matrices, $b$ antisymmetric while $g$ symmetric and positive definite.   The dynamics of classical strings in this background is
defined by the second order Lagrangian
 \be  S[x(\tau,\sigma)]= \!\int  d\tau \oint \left(g_{ij}+b_{ij}\right)\d_+x^i\d_- x^j, \label{282}\ee
 which corresponds to the dynamical system  with  the Darboux symplectic form
 \be \omega=\oint \delta p_i(\sigma)\wedge  \delta x^i(\sigma)\label{291}\ee and with
 the Hamiltonian
\be  H=\jp \int d\sigma g^{ij}\left(p_i-b_{ik}x'^k \right)\left(p_j-b_{jl}x'^l \right) + \jp \int d\sigma g_{ij}x'^ix'^j.\label{286}\ee

\medskip

We now introduce new variables $\tilde p^i,\tilde x_i$ on the phase space by the formulae
\be \tilde p^i:=x'^{i}, \quad \tilde x^{i}=\int p_i.\label{288}\ee
The new variables are also the Darboux ones because it is easy to check that it holds
 \be \omega=\oint \delta \tilde p^i(\sigma)\wedge  \delta \tilde x_i(\sigma).\ee
 Moreover, the Hamiltonian \eqref{286}
can be rewritten in the new variables
as
\be  H=\jp \oint  \tilde g_{ij}\left(\tilde p^i-\tilde b^{ik}x'_k \right)\left(\tilde p_j-\tilde b^{jl}x'_l \right) + \jp \oint \tilde g_{ij}\tilde x'_i\tilde x'_j,\label{294}\ee
which means that it  corresponds to the second order Lagrangian
 \be  S[\tilde x(\tau,\sigma)]= \!\int  d\tau \oint \left(\tilde g^{ij}+\tilde b^{ij}\right)\d_+\tilde x_i\d_- \tilde x_j, \label{296}\ee
 where $\tilde x_i$ are again the angle variables ranging from $0$ to $2\pi$.
Here the dual geometric data $\tilde g$, $\tilde b$ are functions of the original data $g$, $b$ in the sense of the formula
 \be (\tilde g^{ik}+\tilde b^{ik})
 (g_{kj}+b_{kj})=\delta^i_j.\label{298}
 \ee
Said in other words, the  dual geometry
$\tilde g$, $\tilde b$ is obtained by taking the symmetric and the antisymmetric part of the inverse matrix $(g+b)^{-1}$. 

\medskip

The simplest set up is the one-dimensional one
  when the string moves on a single circle $S^1$ parametrized by the angle coordinate $x^1\equiv x$. In this case there is no Kalb-Ramond field and there is one constant  component of the metric tensor ${g_{11}=R}$.
The T-duality then establishes the dynamical equivalence of strings moving on circle backgrounds with flipped radia:
\be S[x ]= R\!\int d\tau \oint\ \! \! \  \! \! \d_+x\d_- x, \quad 
  \tilde S[\tilde x ]=\frac{1}{R} \!\int  d\tau \oint \  \! \!  \  \! \!  \d_+\tilde x\d_- \tilde x.\label{310}\ee

\medskip

\noindent {\bf Remark 2.4:} {\small There are some subtle points related to the canonical transformation \eqref{288} because the prime (standing for the $\partial_\sigma$ derivative) kills the zero mode of the variable $x$ and the integral of $p_i$ is defined up to a constant.  We do not discuss those subtleties here in order to move on rapidly towards the non-Abelian generalizations but the interested reader may consult Ref.\cite{KS97} for a more detailed discussion in the classical and also in the quantum case.}

\subsection{$\E$-model formulation of the Abelian T-duality}
Consider a direct product Lie group $D=U(1)\times U(1)$ and its loop group $LD$ consisting of the set of maps  from the circle ${  S^1}$  into $D$ equipped with the pointwise group multiplication.
 If we parametrize the elements $l$ of $LD$ by the 
phases \be l=(e^{ix}, e^{i\tilde x}),\label{par0}\ee then the
 quantity  \be l'l^{-1}\!=  \ri x' \oplus \ri \tilde x'\ee is  the element of the Lie algebra $L\D$. Because it holds $p=\tilde x'$, we can  
 rewrite the Hamiltonian of the  simplest $ {R\to 1/R}$  example of the Abelian T-duality   as 
\be H=\jp \oint\left(\frac{p^2}{R}+Rx'^2\right) =\jp\oint \bigl(l'l^{-1},\E \ \! l'l^{-1}\bigr)_{\D},\ee
 where the  bilinear form ${ (.,.)_{\D}}$ on the Lie algebra $\D$ and the operator ${\E}:\D\to D$    are defined as
\be  \left(\ri u\oplus\ri v,\ri w\oplus \ri z\!\right)_{\D}=uz+vw,\quad \E(\ri u\oplus \ri v)=R^{-1}\ri v\oplus R\ri u, \quad u,v,w,z\in\br.\label{bif}\ee
Furthermore the Darboux property of the phase space coordinates ${p,x}$ and $\tilde p,\tilde x$ gives the following expression for the symplectic form:
\be   \omega =-\jp\oint \bigl(l^{-1}\delta l\stackrel{\wedge}{,}(l^{-1}\delta l)'\bigr)_{\D}.\label{447}\ee
Indeed,   from Eqs.\eqref{par0},\eqref{bif} and \eqref{288}  we infer
$$  -\jp\oint \bigl(l^{-1}\delta l\stackrel{\wedge}{,}(l^{-1}\delta l)'\bigr)_{\D}=$$\be=\jp \oint \delta p_i(\sigma)\wedge  \delta x^i(\sigma)+\jp \oint \delta \tilde p^i(\sigma)\wedge  \delta \tilde x_i(\sigma)=\oint \delta p_i(\sigma)\wedge  \delta x^i(\sigma)=\oint \delta \tilde p^i(\sigma)\wedge  \delta \tilde x_i(\sigma).\ee

We can play a similar game for the higher-dimensional Abelian T-duality relating the models \eqref{282} and \eqref{296}.
Consider a direct product Lie group $D=U(1)^n\times U(1)^n$ and parametrize the elements $l$ of its loop group $LD$   by the 
phases \be l=(e^{ix^1},\dots,e^{ix^n},  e^{\ri\tilde x_1},\dots, e^{\ri\tilde x_n})=(e^{ x^jT_j},e^{\tilde x_j\tilde T^j}):= (e^{ \boldsymbol{x}},e^{\boldsymbol{\tilde x}}),\ee
where $T_j,\tilde T^j$ form   basis of the Lie algebra $\D$.

Then the
 quantity  $$ l'l^{-1}\!= x'^jT_j \oplus   \tilde x'_j\tilde T^j=   \bx' \oplus   \tbx'$$ is  the element of the Lie algebra $L\D$. Because it holds $p_j=\tilde x'_j$, we can  
 rewrite the Hamiltonian \eqref{286}    of the higher-dimensional  Abelian T-duality   as 
\be H=\jp \oint g^{ij}\left(p_i-b_{ik}x'^k \right)\left(p_j-b_{jl}x'^l \right) + \jp \int d\sigma g_{ij}x'^ix'^j =\jp\oint \bigl(l'l^{-1},\E \ \! l'l^{-1}\bigr)_{\D},\ee
 where the  bilinear form ${ (.,.)_{\D}}$ on the Lie algebra $\D$ and the operator ${\E}:\D\to D$    are defined as
\be  \left(\boldsymbol{u}\oplus\boldsymbol{v},\boldsymbol{w}\oplus\boldsymbol{z}\right)_{\D}=u^jz_j+v_jw^j,\ee
\be \E(\boldsymbol{u}\oplus\boldsymbol{v})=  g^{jk}(v_k -b_{kl}u^l)T_j\oplus \left((g_{jk}-b_{jm}g^{ml}b_{lk})u^k +b_{jm}g^{mk}v_k \right)\tilde T^j.\ee
Furthermore the Darboux property of the phase space coordinates ${p_i,x^i}$ and $\tilde p^i,\tilde x_i$ gives  the following   formula for the symplectic form \eqref{291}:
\be   \omega =-\jp\oint   \bigl(l^{-1}\delta l\stackrel{\wedge}{,}(l^{-1}\delta l)'\bigr)_{\D}.\label{349}\ee
\subsection{Poisson-Lie T-duality}
We have observed in the previous subsection that  the higher-dimensional Abelian T-duality is structurally the same as the one-dimensional one. This means that

\bigskip

\noindent 1) There is an (Abelian) Lie group $D$ of even dimension, the Lie algebra $\D$ of which is equipped with a  non-degenerate symmetric ad-invariant  bilinear form $(.,.)_\D$ of maximally Lorentzian signature $(+\dots +,-\dots -)$.

\medskip
 
 \noindent 2) There are two half-dimensional connected isotropic Lie subgroups $K$ and $\tilde K$ of $D$  where the term "isotropic" means that the restriction of the bilinear form $(.,.)_\D$ on the Lie algebras $\K$ and $\tilde\K$ vanishes ($\K$ and $\tilde\K$  are respectively generated by the basis $T_j$ and $\tilde T^j$).
 
 \medskip
 
 \noindent 3) There is a linear operator $\E:\D\to\D$ which squares to the identity on $\D$, i.e. $\E^2={\rm Id}$, $\E$ is self-adjoint with respect to the bilinear form $(.,.)_\D$, i.e. $(\E .,.)_\D=(.,\E .)_\D$ and, finally, the $\E$-dependent symmetric bilinear form $(.,\E .)_\D$ is strictly positive definite.
 
 \bigskip

 The distinction between the higher-dimensional case and the one-dimensional one is  hidden just in the choice
 of the group $D$. In the higher-dimensional case 
 we have $D=U(1)^n\times U(1)^n$ and in the one-dimensional it is  $D=U(1)\times U(1)$.
 
 \medskip

 The moral of the story is as follows: to the data 1), 2) and 3) there is naturally associated  the so called
 $\E$-model, which is the loop dynamical system living on the phase space $LD$, with the symplectic form 
 \be   \omega =-\jp\oint   \bigl(l^{-1}\delta l\stackrel{\wedge}{,}(l^{-1}\delta l)'\bigr)_{\D}\label{373}\ee
 and with the Hamiltonian
 \be H_\E  =\jp\oint \bigl(l'l^{-1},\E \ \! l'l^{-1}\bigr)_{\D}.\label{375}\ee
 In the case of the Abelian group $D$, we have seen that the $\E$-model is the dynamical system describing the strings moving on the background \eqref{282} but also on the background \eqref{296}.
 
 \medskip
 
 What is the Poisson-Lie T-duality? In the $\E$-model picture, it is   the generalization of the Abelian T-duality in which we just replace the Abelian Lie group $D$ by a non-Abelian one. It is really as simple as this!
Indeed, we are now going to show that taking a non-Abelian Lie group $D$ supplemented with the structures 1), 2) and 3),  the corresponding $\E$-model \eqref{373}, \eqref{375}
 is the dynamical system describing the strings moving on two geometrically different backgrounds.

\medskip

 \noindent {\bf Remark 2.5}: {\small Every Lie group ${D}$ fulfilling the properties 1) and 2)  is referred to as the {  Drinfeld double}.
 The Drinfeld doubles are so abundant that they were not even classified, the Poisson-Lie T-duality constitutes therefore a  genuine factory to produce plenty of examples of the T-dual geometries.}
 
 \medskip
 
  How to extract two mutually dual Riemannian-Kalb-Ramond geometries  from the $\E$-model data $(LD,\omega,H_\E)$? It is particularly simple to achieve this goal
    in a special case where the Drinfeld double is {\it perfect}, which means that $D$ is topologically the direct product $D=K\times \tilde K$ and this direct product decomposition is compatible with the group multiplication.
Said in other words, every element $l$ of the perfect Drinfeld double can be written in an unique way as the group   multiplication (in the sense of the group $D$) of one element of the group $K$ and another of  $\tilde K$:
\be l=k\tilde h,\qquad k\in K, \quad \tilde h\in \tilde K.\label{397}\ee
 Inserting the decomposition \eqref{397} into \eqref{373} and into \eqref{375}, we obtain easily
 \be \omega=\delta\left(\oint (\tilde\Lambda,k^{-1}\delta k)_\D\right),\label{399}\ee
\be H_\E(k,\tilde\Lambda) =\jp \oint (\partial_\sigma kk^{-1}+k\tilde\Lambda k^{-1},\E(\partial_\sigma kk^{-1}+ k\tilde\Lambda k^{-1}))_\D,\label{400}\ee
 where $\tilde\Lambda=\partial_\sigma \tilde h\tilde h^{-1}$ is a $\tilde\K$-valued field playing the role of a generalized momentum.
 
 \medskip
 
 Knowing the symplectic form and the Hamiltonian in the variables $k,\tilde\Lambda$, we can write down the first order action of the corresponding dynamical system as
\be S_\E=\int d\tau \oint  \left( (\tilde\Lambda,k^{-1}\partial_\tau k)_\D-  \jp  (\partial_\sigma kk^{-1}+k\tilde\Lambda k^{-1},\E(\partial_\sigma kk^{-1}+ k\tilde\Lambda k^{-1}))_\D\right). \label{foa}\ee
The quadratic dependence on $\tilde\Lambda$ makes then easy to find the second order action depending solely  on $k\in K$. It reads:  
 \be S_E(k)=\jp\int d\tau \oint  
 \left(\left(E+\Pi(k)\right)^{-1}\partial_+kk^{-1}, \partial_-kk^{-1}\right)_\D.\label{407}\ee
 Here  
  the linear operator $E:\tilde\K\to\K$ is defined by the operator $\E:\D\to\D$ in the way that   its graph $\{\tilde x+E\tilde x,\tilde x\in\tilde\K\}$ coincides with the image of the operator Id$+\E$. As far as
the $k$-dependent operator $\Pi(k):\tilde\K\to\K$ is concerned it is given by  the expression
\be \Pi(k)=-\J{\rm Ad}_k\tilde\J {\rm Ad}_{k^{-1}}\tilde\J.\label{411}\ee
Here Ad$_k$ stands for the adjoint action on $\D$ of the element $k\in K\subset D$ and   $\J,\tilde\J$ are projectors; $\J$ projects to $\K$ with the kernel $\tilde\K$ and  $\tilde\J$ projects to $\tilde\K$ with the kernel $\K$.

\medskip

Starting from the opposite decomposition of the Drinfeld double
\be l=\tilde k h,\qquad \tilde k\in \tilde K, \quad h\in K \label{414}\ee
and repeating the same derivation with the roles of $K$ and $\tilde K$ exchanged, we arrive at the dual  second order action for the dual field $\tilde k$ living in $\tilde K$:
\be \tilde S_{\tilde E}(\tilde k)=\jp\int d\tau \oint  
 \left(\left(\tilde E+\tilde\Pi(\tilde k)\right)^{-1}\partial_+\tilde k\tilde k^{-1}, \partial_-\tilde k\tilde k^{-1}\right)_\D,\label{418}\ee
 where
 \be \tilde\Pi(\tilde k)=-\tilde\J{\rm Ad}_{\tilde k}\J {\rm Ad}_{\tilde k^{-1}}\J.\label{420}\ee
 
 Of course, the linear operators $E:\tilde\K\to\K$ and
 $\tilde E:\K\to\tilde\K$ are not independent objects because their graphs coincide, being equal to the image of the operator
 Id$+\E$. It facts, it turns out that they are inverse to each other.
 
 \medskip
 
 \noindent {\bf Remark 2.6:} {\small The action \eqref{407} as well as the dual action \eqref{418} define the Riemannian-Kalb-Ramond geometries respectively on the group targets $K$ and $\tilde K$. It is not difficult to extract from
 \eqref{407}, \eqref{418} explicit formulas for the metrics and for the Kalb-Ramond fields but we prefer to stick on the elegant coordinate invariant formulations  \eqref{407}, \eqref{418} of those Riemannian-Kalb-Ramond geometries.}
 
 \medskip
 
 \noindent {\bf Remark 2.7:} {\small Remarkably, the operator  $\Pi(k):\tilde\K\to\K$ (as well as its dual $\tilde\Pi(\tilde k):\K\to\tilde\K$) defines a Poisson bracket on the group manifolds $K$ (on $\tilde K$).
 
 \medskip 
 
  \noindent {\bf Remark 2.8:} {\small If one of the subgroups $K$, $\tilde K$ is Abelian and the other is non-Abelian, the Poisson-Lie T-duality reduces to the so called non-Abelian T-duality introduced in Refs.\cite{DQ,FT,FJ}.
  
  \medskip

 In particular, if  $f_1,f_2$ are two functions on the group $K$, their bracket  defined by the  formula:
\be \{f_1,f_2\}_{K}(k)=(\nabla^L f_1,\Pi(k)\nabla^L f_2)_\D\label{435}\ee
turns out to be the Poisson one.  
Here $\nabla^{L}$ is $\tilde\K$-valued differential operator acting on the functions on $K$  as
\be (\nabla^L f, x)_\D(k):= (\nabla^L_{x}f)(k)\equiv \frac{df(e^{sx}k)}{ds}\bigg\vert_{s=0}, \qquad x\in\K.\label{438}\ee
Actually, the Poisson bracket \eqref{435} is of the so called Poisson-Lie type.}

\medskip

An example of the perfect double $D$ was constructed by Lu and Weinstein \cite{LW}  and it is the special complex linear group ${ SL(n,\bc)}$ viewed as {\it real} group (i.e. it has the dimension $2(n^2-1)$ as the real manifold). 
The  non-degenerate symmetric ad-invariant bilinear form ${ (.,.)_{\D}}$ on the Lie algebra ${sl(n,\bc)}$ is defined by taking the imaginary
 part of the trace
 \be {(X,Y)_{\D}= \Im\tr(XY), \quad X,Y\in  sl(n,\bc)}\ee
 and has indeed  the maximally Lorentzian signature ${(n^2-1,n^2-1)}$. The two half-dimensional subgroups $K$ and $ \tilde K$ are
 respectively the special unitary group ${ SU(n)}$ and  the upper-triangular group ${ AN}$ with real positive numbers on the diagonal the product of which is equal to $1$.
 
 \medskip
 
 Every $\E$-model yields a nontrivial T-duality pattern 
 even in the case when the double $D$ is not perfect. In this non-perfect case, the most convenient
 way of casting the actions of the pair of mutually dual $\sigma$-models is in terms of the following expressions
 \begin{subequations}\label{2ndorderactions}\begin{align}S_\E(l)&=\frac{1}{4}\int \delta ^{-1}\oint \biggl(\delta ll^{-1},[\partial_\sigma ll^{-1},\delta ll^{-1}]\biggr)_\D +\label{origaction}\\ &+\frac{1}{4} \int d\tau\oint \biggl( \partial_+ ll^{-1},Q^-_l \partial_- ll^{-1}\biggr)_\D-\frac{1}{4} \int d\tau\oint \biggl(Q^+_l\partial_+ ll^{-1}, \partial_- ll^{-1}\biggr)_\D, \nonumber\\
  \tilde S_\E(l)&=\frac{1}{4}\int \delta^{-1}\oint \biggl(\delta ll^{-1},[\partial_\sigma ll^{-1},\delta ll^{-1}]\biggr)_\D  \nonumber\\ &+\frac{1}{4} \int d\tau\oint \biggl( \partial_+ ll^{-1},\tilde Q^-_l \partial_- ll^{-1}\biggr)_\D-\frac{1}{4} \int d\tau\oint \biggl(\tilde Q^+_l\partial_+ ll^{-1}, \partial_- ll^{-1}\biggr)_\D. \label{dualaction}\end{align}\end{subequations}
 Here $l(\tau,\sigma)\in D$ is a field configuration,
 $\delta^{-1}$ is a (symbolic) inverse of the de Rham differential and $Q^\pm_l$, $\tilde Q^\pm_l:\D\to\D$ are the projectors  fully characterized by their respective images  and kernels
\be \label{projectors}  \mathrm{Im}(Q_l^\pm)=Ad_l(\tilde\K),\quad
  \mathrm{Im}(\tilde Q_l^\pm)=Ad_l(\K), \quad
  \mathrm{Ker}(Q_l^\pm)=\mathrm{Ker}(\tilde Q_l^\pm)=(1\pm\E)\D.\ee
 It may seem, that both the $\si$-model \eqref{origaction} as well as its dual \eqref{dualaction} live on the target $D$ but, actually, the former lives on the space of cosets $D/\tilde K$ while the latter on the space $D/K$. Indeed, this is because
 the  models \eqref{origaction} and \eqref{dualaction}  enjoy respectively the gauge symmetries \begin{subequations}\label{gaugesymmetries}
 \begin{align}\label{origgaugesym} l(\tau,\sigma)&\to l(\tau,\sigma)\tilde h(\tau,\sigma), \quad \tilde h(\tau,\sigma)\in \tilde K,\\ \label{dualgaugesym} 
  l(\tau,\sigma)&\to l(\tau,\sigma)h(\tau,\sigma),\quad h(\tau,\sigma)\in K.\end{align}\end{subequations}
  To verify that these gauge symmetries take indeed place, we need the standard Polyakov-Wiegmann identity
  \be WZ_\D(gh)=WZ_\D(g)+WZ_\D(h)+2\oint\biggl((g^{-1}\delta g,\partial_\sigma hh^{-1})_\D-(\delta hh^{-1},g^{-1}\partial_\sigma g)_\D\biggr),\label{PW}\ee 
  where
  $$WZ_\D(g):=\delta^{-1}\oint(\delta gg^{-1}\stackrel{\wedge}{,}
[\partial_\sigma gg^{-1},\delta gg^{-1}])_\D$$

 In the case of the perfect double $D$, there is a natural global gauge fixing $l=k$, $k\in K$ which transforms the expression \eqref{origaction} into the action \eqref{407}, while in the dual case the natural gauge fixing is $l=\tilde k$, $\tilde k\in K$   which transforms the expression \eqref{dualaction} into the action \eqref{418}.
 
 \subsection{Dressing cosets}
 There exists a natural generalization of the
 Poisson-Lie T-duality known under the name of the
 dressing coset T-duality \cite{KS97}.  The basic ingredients of this more general duality is again the Drinfeld double $D$,  its half-dimensional maximally isotropic Lie subgroups $K$, $\tilde K$ and the self-adjoint operator $\E:\D\to\D$ squaring to the identity, but newly  we need a (strictly less than half-dimensional) isotropic Lie subgroup $F\in D$ such that the adjoint action of $F$ on $\D$ commutes with $\E$ and the bilinear
 form $(.,\E.)_\D$ restricted to the Lie subalgebra $\F\subset\D$ is non-degenerate. The actions of the mutually dual $\sigma$-models are then again  given by the expressions \eqref{2ndorderactions}, but now the projectors   $Q^\pm_l$, $\tilde Q^\pm_l:\D\to\D$ are determined by more general requirements
\be \label{dressprojectors}  \mathrm{Im}(Q_l^\pm)=Ad_l(\tilde\K),\quad
  \mathrm{Im}(\tilde Q_l^\pm)=Ad_l(\K), \quad 
  \mathrm{Ker}(Q_l^\pm)=\mathrm{Ker}(\tilde Q_l^\pm)=\E_\pm \oplus\F,\ee
 where
 \be \E_\pm\equiv (1\pm\E)\D\cap (\F\oplus\E\F)^\perp.\ee
The properties \eqref{dressprojectors} of the projectors  $Q^\pm_l$, $\tilde Q^\pm_l$ guarantee
that the respective gauge symmetries \eqref{gaugesymmetries} of the $\sigma$-models 
\eqref{2ndorderactions} get augmented by one more common gauge symmetry 
\be l(\tau,\sigma)\to f(\tau,\sigma)l(\tau,\sigma),\quad f(\tau,\sigma)\in F.\ee
This means that the $\sigma$-model \eqref{origaction}   lives on the space of double cosets $F\backslash D/\tilde K$ while the dual $\sigma$-model \eqref{dualaction} lives on the  space $F\backslash D/K$. If the subgroup $F$ is trivial, the dressing coset T-duality reduces to the Poisson-Lie T-duality of the previous section.

\medskip

What is the common symplectic structure  of the dual pair of the dressing coset models? It is given by the symplectic reduction of the symplectic form \eqref{373}   with respect to the  left action of the loop group $LF$ on the loop group $LD$. This action is generated by the moment maps of the form $(l'l^{-1},x)_\D$, $x\in\F$ therefore the reduction is defined by the following constraints imposed on the unreduced phase space $LD$
\be (l'l^{-1},\F)_\D=0.\label{cons}\ee
The space of the orbits of the loop group $LF$ on the constraint surface \eqref{cons} is the reduced  phase space or, equivalently, the phase space of the dressing coset $\sigma$-models. 

\medskip

In order to express the Hamiltonian $H_{dc}$ of the dressing coset $\sigma$-models, we need first to decompose the Lie algebra $\D$ as the following direct sum of vector spaces
\be \D=(\F\oplus \E\F)^\perp\oplus (\F\oplus \E\F).\label{decom}\ee
Given $y\in\D$, we denote by $y_0$ the first term in the decomposition \eqref{decom}. Then the Hamiltonian is given by
\be H_{dc}=\jp\oint ((l'l^{-1})_0,\E(l'l^{-1})_0)_\D.\ee

 \section{Integrability and the Ruijsenaars duality}
  The Ruijsenaars duality connects the duality story with that of integrability.  It takes place in  the context of the dynamics of a finite number of  degrees of freedom, where the notion of integrability
is by now perfectly understood in the mathematical literature.  On the other hand, in the physical literature there can be  sometimes found semi-rigorous formulations (even at the textbook level) which often miss the point what the integrability really is. In particular, those inexact formulations do not emphasize or even mention the crucial issue of \textit{completeness} of the commuting evolution flows generated by the Hamiltonians in involution.  As the result, those imprecise  definitions of integrability are quite empty in the sense that they suit too many dynamical systems and do not guarantee  the special properties of the truly integrable systems. 

 In what follows, we first provide   clarifications of the concepts of completeness and polarization, before  giving the precised definition of the integrability
 and of the Ruijsenaars duality. 
 
 \subsection{Completeness of dynamical systems}
 
 Given a symplectic manifold $P$ and a dynamical system $(P,H)$ on it, the evolution flow $u_t$ generated by the Hamiltonian $H$ is called \textit{complete}
 if it  can be smoothly prolonged to both forward and backward infinities $t\to\pm\infty$. The point $u_0\in P$ is called the \textit{origin} of the flow.
 
 \medskip
 
  \textit{The complete} dynamical system  $(P,H)$  is such  that  all evolution flows on the symplectic manifold $P$, whatever are their origins,  are   complete.

\medskip
It may seem that 
  that the notion of the complete dynamical system is quite difficult to handle in practice because the detailed knowledge of all evolution flows appears to be needed to determine whether a given Hamiltonian does yield  a complete dynamics or not.
However, sometimes shortcut arguments may help to settle the issue of completeness even without having at hand  a detailed description of the flows. For example, if the symplectic manifold $P$ is compact then all flows are complete whatever is the smooth Hamiltonian generating the flows.

\medskip

As a non-compact example of the completeness of the flows, consider the manifold
\be P_n=\{(q^1,\dots,q^n,p_1,\dots,p_n)\in \br^{2n}, \quad q^1>\dots >q^n\}\label{524}\ee
equipped with a symplectic form
\be \omega=dp_j\wedge dq^j.\label{526}\ee
For the Hamiltonian, we take the Calogero-Moser one
 \be H_{\rm CM}(p_j,q^j;g)=\jp\sum_{j=1}^n p_j^2+ \jp\sum_{j\neq k}^n\frac{g^2}{(q^j-q^k)^2},\label{488}\ee
 where $g$ is a coupling constant. 
 
 \medskip

 Since the energy $E_{\rm CM}(u_0)$ on any  flow $u_t\in P_n$ is conserved and the  Calogero-Moser Hamiltonian is given by the positive fonction, the  momenta as well as the velocities of the Calogero-Moser particles are all bounded on the totality of the given flow $u_t$, therefore the particles cannot escape into infinity in a finite time. At the same time, the particles with positions $q_k(t)$ and $q_{k+1}(t)$ always avoid the
singular points $q_k=q_{k+1}$ (in a finite or in an infinite time) again because of the positivity of the Calogero-Moser Hamiltonian and because of  the conservation of energy. The Calogero-Moser dynamical system $(P_n,H_{\rm CM})$ is therefore complete.  

\medskip

On the other hand, it  has to be stressed that    dynamical systems with  non-complete evolution flows are quite frequent and easy to construct. We may e.g.  consider some complete Calogero-Moser flow passing through a point $u\in P_n$  and pick some open neighbourhood $O_u$ of $u$ which does not contain the totality of the flow. The open neighbourhood  
$O_u$ is itself a symplectic manifold and the same Hamiltonian $H_{\rm CM}$ which generates the complete flow on $P_n$ will yield a non-complete flow on $O_u$. The dynamical system $(O_u,H_{\rm CM})$ is therefore non-complete.

\subsection{Polarizations}
Denote by Fun$(P)$ the commutative and associative algebra with respect to the point-wise multiplication which consists of all smooth functions on a symplectic manifold $P$. A  \textit{polarization} of $P$ is any maximal Poisson commuting subalgebra
$\A\subset\fp$. Said in other words,  $\A$ is the commutative and associative subalgebra of $\fp$ such that the Poisson bracket $\{f_1,f_2\}$  vanishes whenever $f_1,f_2\in\A$. Moreover, $\A$  is not strictly contained in any  subalgebra of $\fp$ having the same property.

\medskip 

Before advancing further, it is perhaps worth making a connection
of our definition of the polarization
with the one usually given, 
say, in the context of geometric quantization, where the polarization is understood as the smooth   distribution of   Lagrangian subspaces of the tangent spaces of the symplectic manifold. In fact, given the polarization $\A$, those Lagrangian subspaces are spanned by the Hamiltonian vector fields of the form $\{a,.\}$, $a\in \A$, where $\{.,.\}$ denotes the Poisson bracket corresponding to the symplectic form.

\medskip

The concept of polarization is not only important for the geometric quantization but it plays an essential role
also in the study of purely classical integrable systems. Actually, this classical role is even two-fold. First of all, the polarization is necessary for the physical interpretation of the flows:

\medskip

\textit{The dynamical system with physical interpretation} is a triple $(P,\A,H)$, where $P$ is a symplectic manifold, $\A$ is a polarization and $H$ is a Hamiltonian. The polarization $\A$ bears  the name "physical position polarization".

\medskip

\medskip

  The principal reason for the terminology   "physical interpretation" or "physical position polarization" comes from the fact that for physical applications it is often necessary to distinguish which coordinates on the phase space $P$ correspond to physical positions of particles and which to momenta. If the phase space has the topology ${\mathbb R}^{2n}$ and possesses a global Darboux parametrization $(p_j,q^j)\in\br^{2n}$, then the set of all functions $f(q^j)$ on $\br^{2n}$, which are independent on $p_j$, constitutes the physical position polarization $\A$. This polarization subalgebra is  generated by the global coordinates $q^j$ which are functionally independent because everywhere on $\br^{2n}$  it holds
\be dq^1\wedge dq^2\wedge \dots \wedge dq^n\neq 0.\ee

If the space of the physical positions is some topologically nontrivial
  manifold $M$ (called  the configuration space), the phase space
  is the cotangent bundle $T^*M$. In this case, the effective separation of the "position" and the "momenta" can still be done via the concept of the polarization, although no global  Darboux coordinates exist and neither do exist global position coordinates $q^j$ which would generate the subalgebra $\A$. Remarkably, the easiest  way to proceed in this topologically non-trivial case  does not consist in covering $M$ by some local coordinate patches but it remains  global and conceptual. If we denote by $\pi$ the canonical projection $\pi:T^*M\to M$, then
the  physical position polarization $\A$ on $T^*M$ consists of all functions on $T^*M$ which are the pull-backs $\pi^*f$ of functions $f$ on $M$. For completeness, we mention that the symplectic form $\omega$ on $T^*M$ is given by
\be \omega=d\alpha,\ee
where $\alpha$ is a $1$-form on $T^*M$ defined at a point $(x,\beta)\in T^*M$ as
\be \alpha\vert_{(x,\beta)}=\pi^*\beta,  \qquad  x\in M, \quad \beta\in T^*_x M.\ee

In the case of  the Calogero-Moser dynamical system
$(P_n,H_{\rm CM})$,   the things do not require  the level of sophistication described in the previous paragraph, since the global position coordinates $q^j$ exist albeit they are restricted by the inequalities $q^1>q^2>\dots>q^n$. The physical position polarization $\A$ is therefore the subalgebra of functions on $P_n$ which do not depend on the momenta $p_j$:
\be \A=\{f\in {\rm Fun}(P_n), \quad \d_{p_j}f=0, \quad j=1,\dots, n\}.\label{549}\ee

The polarization subalgebra $\A$ defined in this way  stands simply for the coordinate-invariant way of introducing  the physical configuration space in which the Calogero-Moser particles move. 

\subsection{Integrability}

Now we describe the second role played by the concept of the polarization in the story of the classical integrability.

\medskip

A polarization $\B$ is called {\it complete} if for every Hamiltonian $ H\in \B$ the dynamical system $(P, H)$ is complete.

\medskip

{\it The integrable dynamical system} $(P,\B)$ is any complete polarization $\B$ on the symplectic manifold $P$. The complete polarization $\B$ bears  the name "the action variables polarization".

\medskip

In practice, the integrable dynamical system  
$(P,\B)$ is often  constructed starting from a given complete dynamical system $(P,H)$ (like the Calogero-Moser one)  by finding the complete polarization subalgebra $\B$ which contains the Hamiltonian $H$ as its element.  Since the elements of $\B$ do Poisson-commute among themselves,   they all commute in particular with $H$ and thus constitute maximal set of conserved integrals of motion in involution. 

\medskip

It is perhaps worth remarking in this respect, that many authors define  the integrability 
in a similar way, that is by claiming that  the maximal set of the conserved integrals of motion must generate a polarization subalgebra. However,  this claim is not very restrictive if the    completeness of this polarization is not required.    In reality, it is the requirement of the completeness   which makes the integrable systems so rare and interesting. In particular, as it was pointed out in Ref.\cite{RuiBanff},    the completeness of the polarization
 guarantees the validity of the Liouville-Arnold theorem \cite{L,A}  which says the following: 
 If a dynamical system $(P,\B)$ is integrable then its phase space $P$ can be decomposed into open, connected, pair-wise non-intersecting, flow-invariant parts $U_\alpha\subset P$ such that $P=\cup_\alpha\bar U_\alpha$ and on each 
part $U_\alpha$ there can be introduced the so-called
{\it action-angle} variables $I_j,\theta^j$.  Those special variables verify few important properties. First of all, the action variables $I_\alpha$ are restrictions to $U_\alpha$ of suitable elements of the complete polarization $\B$, they are functionally independent and the restriction of the symplectic form $\omega$ on  
$U_\alpha$ takes the form
\be \omega\vert_{U_\alpha}=dI_j\wedge d\theta^j.\ee
Moreover, the Hamiltonian $H$ restricted onto $U_\alpha$ depends exclusively on the action variables $I_j$ and does not depend on the 
angles $\theta^j$. It follows that the dynamics of the system becomes explicitly determined: the action variables $I_j$
are constants of motion and the evolution flows of the variables
$\theta^j$ are complete on $U_\alpha$ and linear in time. Actually, if the phase space $P$ is non-compact some of the thetas need not be angles and they can be non-compact, too.

\medskip

It should be finally pointed out that the existence of a complete
polarization $\B$ containing  the given    complete Hamiltonian $H$ is by no means granted. Many dynamical systems
$(P,H)$ are not integrable and the difficult  problem is to find out for which     Hamiltonian $H$ it exists the complete polarization $\B$ which contains it and for which one in turn such a polarization  does not exist.

\subsection{Ruijsenaars duality}

{\it An integrable dynamical system with physical interpretation} is a triple $(P,\A,\B)$, where  $\A$, $\B$  are the polarizations on the symplectic manifold $P$; the physical position polarization  $\A$ may be arbitrary but the action variables polarization $\B$ must be complete.

\medskip 

Very often, however, the physical position polarization $\A$ turns out to be complete, too. In particular,  this is  always the case when the phase space $P$ has the structure of the cotangent bundle $T^*M$ and  $\A=\pi^*{\rm Fun}(M)$.
In any case, if the polarization $ \A$ is complete then, mathematically speaking, it appears in the triple $(P,\A,\B)$ on the same footing as the complete polarization $\B$.  The   Ruijsenaars duality then flips the   roles of the polarizations $\A$ and $\B$, so, from the dual point of view, it is not $\A$ which provides  the physical interpretation to the integrable system $(P,\B)$ but it is $\B$ which furnishes   the integrable system $(P,\A)$ with the physical interpretation.   The {\it Ruijsenaars dual} of the integrable dynamical system $(P, \A,\B)$ with physical interpretation is 
thus the integrable  dynamical system  $(P,\B,\A)$ with physical interpretation.
 
 \medskip
 
 In practice, the study of the Ruijsenaars duality concerns   certain deformations of the complete dynamical systems of the Calogero-Moser type.  
 One typically starts with the study of  such complete deformed dynamical system with physical interpretation $(P,\A,H(\bos{c}))$, where $\bos{c}=(c^1,\dots,c^s)$ is the set of the  deformation parameters. The first task consists in looking for the complete polarization $\B$
 containing $H(\bos{c})$. If $\B$ can be found, then the second task consists in looking for a particular element $\tilde H(\bos{c})\in\A$ which would have either the same or, possibly, another deformed Calogero-Moser form  when expressed in the (original point of view) action-angle variables.
 
 \medskip
 
 We now illustrate the phenomenon of the Ruijsenaars duality on two examples for the simplest non-trivial case
 $n=2$. We shall study the case of arbitrary $n$ in the next section.
 
 \medskip
 
 We start with the non-deformed Calogero-Moser Hamiltonian
  \be H_{\rm CM}(p_1,p_2,q^1,q^2;g)=\jp p_1^2+\jp p_2^2+  \frac{g^2}{(q^1-q^2)^2}\label{622}\ee
  defined on the phase space
 \be P_{2}=\{(q^1,q^2,p_1,p_2)\in \br^{4}, \quad q^1> q^2\}.\label{625}\ee
The manifold $P_2$ is equipped with a symplectic form
\be \omega=dp_1\wedge dq^1+dp_2\wedge dq^2\label{627}\ee
giving rise to the Poisson bracket of functions $a,b$ on $P_2$
\be \{a,b\}_2=\d_{q_1}a\d_{p^1}b- \d_{p_1}b\d_{q^1}a
+\d_{q_2}a\d_{p^2}b- \d_{p_2}b\d_{q^2}a.\ee
Consider a map $\rho_2: P_2\to P_2$ defined by
 \begin{subequations}  \label{806}
 \begin{align}
 \tilde q_1&=\jp\left(\frac{4g^2}{(q^1-q^2)^2}+ (p_1-p_2)^2\right)^{\jp}-\jp(p_1+p_2),\\ \tilde q_2&=-\jp\left(\frac{4g^2}{(q^1-q^2)^2}+(p_1-p_2)^2\right)^{\jp}-\jp(p_1+p_2), \\ \tilde p^1&=-\jp(p_1-p_2)(q^1-q^2)\left(\frac{4g^2}{(q^1-q^2)^2}+(p_1-p_2)^2\right)^{-\jp}+\jp(q^1+q^2),  \\ \tilde p^2&=\jp(p_1-p_2)(q^1-q^2)\left(\frac{4g^2}{(q^1-q^2)^2}+(p_1-p_2)^2\right)^{-\jp}+\jp(q^1+q^2).   \end{align}\end{subequations}
The map $\rho_2$ is involutive, which means that $\rho_2\circ\rho_2={\rm Id}$. As such, $\rho_2$ is bijective and, in fact, it is the symplectomorphism of the phase space $P_2$, because it is easy to establish that it holds
\be \omega_2=dp_1\wedge dq^1+dp_2\wedge dq^2=d\tilde p^1\wedge d\tilde q_1+d\tilde p^2\wedge d\tilde q_2.\label{634}\ee
Recall that the physical position polarization $\A_2$ of the $n=2$ Calogero-Moser system is defined as
\be \A_2=\{f\in {\rm Fun}(P_2), \quad \d_{p_1}f=\d_{p_2}f=0\}.\label{638}\ee
Is this polarization   complete?  To answer affirmatively this question, we observe that every element $f(q^1,q^2)\in\tilde\A_2$ generates the flows $u_t$ linear in time
\be q^j(u_t)=q^j(u_0), \quad   p_j(u_t)=p_j(u_0)-(\d_{q^j}f)(q^1(u_0),q^2(u_0))t, \qquad j=1,2,\label{643}\ee
all flows generated by given $f(q^1,q^2)\in\A_2$ are therefore obviously complete.

\medskip

Define  
\be \B_2=\rho_2^*\A_2=\{f\in{\rm Fun}(P_2); f=\tilde f\circ \rho_2 \quad {\rm for}\quad \tilde f\in\A_2\}\ee
or, equivalently,
\be \B_2=\{f\in {\rm Fun}(P_2), \quad \d_{\tilde p_1}f=\d_{\tilde p_2}f=0\}.\label{639}\ee
Being the pull-back of the complete polarization $\A_2$ by the symplectomorphism, $\B_2$  is also the complete polarization. Moreover, the Calogero-Moser Hamiltonian $H_{\rm CM}$ belongs to $\B_2$, because it holds
\be H_{\rm CM} =\jp p_1^2+\jp p_2^2+  \frac{g^2}{(q^1-q^2)^2} =\jp \tilde q_1^2+\jp \tilde q_2^2.\ee
We thus succeeded to embed the complete $n=2$ Calogero-Moser dynamical system $(P_2,H_{\rm CM})$ into the integrable system $(P_2,\B_2)$, proving in this way its integrability. Of course, we have even more than that at hand, since we have
also the physical interpretation $(P_2,\A_2,\B_2)$ of this dynamical system.

\medskip

What about the Ruijsenaars dual $(P_2, \B_2,\A_2)$? Is there some dual Hamiltonian $\tilde H\in\A_2$ which would have (possibly deformed) Calogero-Moser form
in the  action-angle variables $\tilde q_j,\tilde p^j$? Obviously 
there is a one because of the involutivity of the symplectomorphism $\rho_2$! Indeed we set 
\be \tilde H=\jp (q^1)^2+\jp (q^2)^2= \jp \tilde p_1^2+\jp \tilde p_2^2+\frac{g^2}{(\tilde q_1-\tilde q_2)^2}.\ee
The existence of the dual Calogero-Moser Hamiltonian is usually interpreted by saying that  the Ruijsenaars dual of the $n=2$ Calogero-Moser
dynamical system is the same $n=2$  Calogero-Moser dynamical system.

\medskip

Our second elementary example of the Ruijsenaars duality is slightly more involved, although it still concerns   the
same four-dimensional phase space $P_2$ \eqref{625} equipped with the symplectic form $\omega_2$ \eqref{627}. However, we consider now the following hyperbolic Sutherland deformation of the Calogero-Moser Hamiltonian
\be  H_{\rm hypS}(c;p_1,p_1,q^1,q^2;g)=\jp p_1^2+\jp p_2^2+  \frac{g^2}{4c^2\sinh^2{\frac{q^1-q^2}{2c}}}.\ee
Note that it holds
\be \lim_{c\to\infty}H_{\rm hypS}(c;p_1,p_1,q^1,q^2;g)=H_{\rm CM}(p_1,p_2,q^1,q^2;g),\ee
which means that the $n=2$ hyperbolic Sutherland dynamical system
$(P_2, H_{\rm hypS})$ is the one-parametric  deformation of the $n=2$ Calogero-Moser system $(P_2, H_{\rm CM})$.

\medskip

Is  $n=2$ hyperbolic Sutherland model $(P_2, H_{\rm hypS})$  integrable? Or, said differently, does it exist an integrable model $(P_2,\B)$ such that $H_{\rm hypS}\in\B$? Yes, it does.  To prove it, we construct an appropriate symplectomorphism $\rho_{c,2}: P_2\to P_2$ such that
the pull-back $\rho^*_{c,2}\A_2$ of the complete (physical position) polarization $\A_2$ \eqref{638} will be the complete polarization containing the hyperbolic Sutherland Hamiltonian $H_{\rm hypS}$.

\medskip

Consider a point-canonical transformation 
$\alpha_2: (q^1,q^2,p_1,p_2)\to (x^1,x^2,\pi_1,\pi_2)$ defined by
\be x^1=\jp (q^1+q^2)+c\sinh{\frac{q^1-q^2}{2c}}, \quad x^2=\jp (q^1+q^2)-c\sinh{\frac{q^1-q^2}{2c}},\ee
\be \pi_1=\jp(p_1+p_2)+\frac{p_1-p_2}{2\cosh{\frac{q^1-q^2}{2c}}}, \quad \pi_2=\jp(p_1+p_2)-\frac{p_1-p_2}{2\cosh{\frac{q^1-q^2}{2c}}} \ee
and also another point-canonical transformation $\tilde\alpha_2:(\tilde q_1,\tilde q_2,\tilde p^1,\tilde p^2)\to (\tilde x_1,\tilde x_2, \tilde\pi^1,\tilde\pi^2)$ defined by
\be \tilde x_1=\jp (\tilde q_1+\tilde q_2)+\frac{\tilde q_1-\tilde q_2}{2\cosh{\frac{\tilde p^1-\tilde p^2}{2c}}}, \quad\tilde x_2=\jp (\tilde q_1+\tilde q_2)-\frac{\tilde q_1-\tilde q_2}{2\cosh{\frac{\tilde p^1-\tilde p^2}{2c}}},\ee
\be \tilde \pi^1=\jp (\tilde p^1+\tilde p^2)+c\sinh{\frac{\tilde p^1-\tilde p^2}{2c}}, \quad \tilde \pi^2=\jp (\tilde p^1+\tilde p^2)-c\sinh{\frac{\tilde p^1-\tilde p^2}{2c}}.\ee
Finally, recall the CM action-angle symplectomorphism  $\rho_2: (x^1,x^2,\pi_1,\pi_2)\to (\tilde x_1,\tilde x_2,\tilde\pi^1,\tilde\pi^2)$ given by Eqs.\eqref{806}
 \begin{subequations}  \label{863}
 \begin{align}
\tilde x_1&=\jp\left(\frac{4g^2}{(x^1-x^2)^2}+ (\pi_1-\pi_2)^2\right)^{\jp}-\jp(\pi_1+\pi_2),\\
\tilde x_2&=-\jp\left(\frac{4g^2}{(x^1-x^2)^2}+ (\pi_1-\pi_2)^2\right)^{\jp}-\jp(\pi_1+\pi_2),\\ \tilde \pi^1&=-\jp(\pi_1-\pi_2)(x^1-x^2)\left(\frac{4g^2}{(x^1-x^2)^2}+(\pi_1-\pi_2)^2\right)^{-\jp}+\jp(x^1+x^2),  \\ \tilde \pi^2&=\jp(\pi_1-\pi_2)(x^1-x^2)\left(\frac{4g^2}{(x^1-x^2)^2}+(\pi_1-\pi_2)^2\right)^{-\jp}+\jp(x^1+x^2).   \end{align}\end{subequations}

\medskip

Define the symplectomorphism $\rho_{c,2}:P_2\to P_2$
by
\be \rho_{c,2}=\tilde\alpha_2^{-1}\circ \rho_2\circ \alpha_2\ee
which correspond to the composition of the
canonical transformations $(q^j,p_j)\to (x^j,\pi_j)\to (\tilde x_j,\tilde\pi^j)\to (\tilde q_j,\tilde p^j)$.
Then we define the complete polarization $\B_{c,2}$ par 
\be \B_{c,2}=\rho_{c,2}^*\A_2\ee
which means that the generators of $\A_{c,2}$ are $\tilde q_j=\rho_{c,2}^* q^j$. The hyperbolic Sutherland Hamiltonian belongs to the complete polarization $\B_{c,2}$ because it holds
  \be H_{\rm hypS}(c;p_1,p_2,q^1,q^2;g)=\jp  p_1^2+ \jp  p_1^2+ \frac{g^2}{4c^2\sinh^2{\frac{q^1-q^2}{2c}}}=\jp\tilde q_1^2+\jp \tilde q_2^2. \label{705}\ee
  
So far we have established the integrability of the   $n=2$ hyperbolic Sutherland model $(P_2,H_{\rm hypS})$ by constructing the integrable system 
$(P_2,\B_{c,2})$ such that
$H_{\rm hypS}\in \B_{c,2}$. If we include the physical interpretation in the game, what would be  the
Ruijsenaars dual   $(P_2,\B_{c,2},\A_2)$ of the hyperbolic Sutherland model  $(P_2,\A_2,\B_{c,2})$?
Is there some dual Hamiltonian $\tilde H\in\A_2$ which would have a deformed Calogero-Moser form in the action-angle variables $(\tilde q_j,\tilde p^j)$, $\tilde q_j\in\B_{c,2}$?  Yes, there is. It is given by the formula \cite{RS}
  $$ \tilde H=c^2\left(\cosh{\frac{q^1}{c}}+\cosh{\frac{q^2}{c}}\right)-2c^2=$$\be =c^2\left(\cosh{\frac{\tilde p^1}{c}}+\cosh{\frac{\tilde p^2}{c}}\right)\left(1+\frac{g^2}{c^2(\tilde q_1-\tilde q_2)^2}\right)^{\frac{1}{2}}-2c^2 =\tilde H_{\rm ratRS}(c;\tilde p^1,\tilde p^2,\tilde q_1,\tilde q_2;g).\label{713}\ee
  Indeed, it is easy to verify that it holds
  \be \lim_{c\to\infty}\tilde H_{\rm ratRS}(c;\tilde p^1,\tilde p^2,\tilde q_1,\tilde q_2;g)=  \jp \tilde p_1^2+\jp \tilde p_2^2+\frac{g^2}{(\tilde q_1-\tilde q_2)^2}=H_{\rm CM}(\tilde p^1,\tilde p^2,\tilde q_1,\tilde q_2;g).\ee
  
  \medskip
 
 \noindent {\bf Remark 3.1:} {\small The  form of the Ruijsenaars-Schneider Hamiltonian \eqref{713} in the tilded variables may surprise some readers because of the non-separation of the kinetic and the potential energy. Moreover, even in the case of the vanishing coupling constant $g$, the "free" Hamiltonian $H_{\rm ratRS}$ has non-polynomial dependence on the momenta.  Nevertheless, such Hamiltonians
 naturally appear e.g. in the description of the dynamics of the  multisolitons in the  sine-Gordon  theory \cite{RS}.}
  
  \medskip
  
It turns out that the Ruijsenaars self-duality of the Calogero-Moser model $(P_n,H_{CM})$ as well as the duality between the hyperbolic Sutherland and rational Ruijsenaars-Schneider models takes place 
also in the case of arbitrary $n$ \cite{siru,FK}. The corresponding Hamiltonians then read
 \be H_{\rm hypS}(c;p_j,q^j;g)=\jp \sum_{j=1}^n p_j^2+\jp\sum_{j\neq k}^n \frac{g^2}{4c^2\sinh^2{\frac{q^j-q^k}{2c}}}, \label{515}\ee
  \be H_{\rm ratRS}(c;\tilde p^j,\tilde q_j;g)=c^2\sum_{j=1}^n\cosh{\left(\frac{\tilde p^j}{c}\right)}\prod_{k\neq j}\sqrt{1+\frac{g^2}{c^2(\tilde q_k-\tilde q_j)^2}}-nc^2.\label{516}\ee
  
  \medskip

 \noindent {\bf Remark 3.2:} {\small It is perhaps worth noting that further deformation of the duality pattern \eqref{515} and \eqref{516} restore the self-duality of the Calogero-Moser system \cite{siru}. Namely, replacing in the expression \eqref{516} the rational functions
 of the dual coordinates $\tilde q_j$ by appropriate hyperbolic functions gives the Hamiltonian of the self-dual hyperbolic Ruijsenaars-Schneider model. }
  
  \medskip
  
  Before proving the  Ruijsenaars self-duality of the Calogero-Moser system for  arbitrary $n$, we  adapt the Ruijsenaars duality in our general schema of dualities outlined at the beginning of Section 2. Thus the
  "structure" is a symplectic manifold $P$ equipped with two complete polarizations
  $\A$ and $\B$. The "meaningful point of view" is any choice of the local Darboux coordinates $p_j,q^j$ such that $q^j$ are the local restrictions of the functions from one of the two polarizations and there is   an element from the other polarization which    looks in the coordinates $q^j,p_j$ like the deformed  Calogero-Moser Hamiltonian.

\medskip

  \subsection{Ruijsenaars self-duality for arbitrary $n$}
  In this section, we prove the Ruijsenaars self-duality of the Calogero-Moser system  for arbitrary $n$ \cite{siru}. Recall
  that the phase space of this system is the manifold $P_n$ defined as 
\be P_n=\{(q^1,\dots,q^n,p_1,\dots,p_n)\in \br^{2n}, \quad q^1>\dots >q^n\}\label{740}\ee
equipped with a symplectic form
\be \omega=dp_j\wedge dq^j\label{742}\ee
and the Hamiltonian is given by
 \be H_{\rm CM}(p_j,q^j;g)=\jp\sum_{j=1}^n p_j^2+ \jp\sum_{j\neq k}^n\frac{g^2}{(q^j-q^k)^2}.\label{744}\ee
The physical position polarization $\A_n$ is defined as
\be \A_n=\{f\in {\rm Fun}(P_n), \quad \d_{p_j}f=0 , \quad j=1,\dots,n\}.\label{746}\ee
The completeness of $\A_n$ can be easily shown by the generalization of the $n=2$ argument \eqref{643}.

\medskip

In the case $n=2$, we have obtained the complete polarization $\B_2$ containing the Calogero-Moser Hamiltonian as the pull-back
$\rho_2^*\A_2$ of the complete physical position polarization $\A_2$ by the symplectomorphism $\rho_2:P_2\to P_2$ given by Eqs.\eqref{806}. In the case
of arbitrary $n$, the complete polarization $\B_n$ containing the Calogero-Moser Hamiltonian is
also obtained as the pullback $\rho_n^*\A_n$ , however
the description of the symplectomorphism $\rho_n:P_n\to P_n$ is known only in terms of implicit functions. Concretely,  
 $\rho_n: (p_j,q^j)\to(\tilde p^j,\tilde q_j)$ is defined by the relations
 \be q^j=\lambda^j(\tilde L),\quad \tilde q_j=-\lambda_j(L), \quad j=1,\dots,n\label{813}\ee
 where $\lambda^j(\tilde L)$ and $\lambda_j(L)$    are respectively  the ordered eigenvalues 
of the matrices $\tilde L$ and $L$ with the matrix elements given by 
\be  L_{jk}= p_j\delta_{jk} -(1-\delta_{jk})\frac{\ri g}{q^j-q^k}, \label{769}\ee
  \be  \tilde L_{jk}= \tilde p^j\delta_{jk} -(1-\delta_{jk})\frac{\ri g}{\tilde q_j-\tilde q_k}.\label{804}\ee
We infer from Eq.\eqref{813}
\be H_{\rm CM}(p_j,q^j;g)=\jp\sum_{j=1}^n p_j^2+ \jp\sum_{j\neq k}^n\frac{g^2}{(q^j-q^k)^2}=\jp\tr L^2
  = \jp\sum_{j=1}^n\tilde q_j^2\in\A_n\label{764}\ee 
  and also 
\be H_{\rm CM}(\tilde p^j,\tilde q_j;g)=\jp\sum_{j=1}^n (\tilde p^j)^2+ \jp\sum_{j\neq k}^n\frac{g^2}{(\tilde q_j-\tilde q_k)^2}=\jp\tr \tilde L^2
  = \jp\sum_{j=1}^n  (q^j)^2\in\tilde \A_n.\label{767}\ee
  Provided that Eqs.\eqref{813} indeed defines a symplectomorphism, we conclude from Eq. \eqref{764} that the Calogero-Moser system is integrable for arbitrary $n$ and,
  from Eq. \eqref{767}, that the Calogero-Moser system is self-dual.
  
  \medskip
  
  We now prove the canonicity of the transformation \eqref{813}
by a nice argument due to Kazhdan, Kostant and Sternberg \cite{KKS}.

\medskip

   Let   $v\in\br^n$ be a vector having all components $v_1,\dots, v_n$ equal to $1$ and let us look for two  Hermitian $n\times n$ matrices  
  $X,Y$ verifying the condition
\be XY-YX=\ri g(\1-v\otimes v^\dagger).\label{765}\ee
The   Hermitian matrix $X$ can be always diagonalized by an unitary matrix $k$, therefore
we can write $X,Y$ as $X=kDk^\dagger$, $Y=kLk^\dagger$, where $D$ is the diagonal matrix
\be D={\rm diag} (q^1,...,q^n), \quad q^1\geq q^2\geq ...\geq q^n.\label{766}\ee
 Eq. \eqref{765} then tells us three things: 
 
 \medskip
 
 1) It holds $q^1> q^2>...>q^n$.
 
 \medskip
 
 2)   
  The unitary matrix $k$ has $v$ as its eigenvector.
  
  \medskip
  
  3) $L$  is the matrix of the form
\be  L_{jk}= p_j\delta_{jk} -(1-\delta_{jk})\frac{\ri g}{q^j-q^k}. \label{769bis}\ee
  
  \medskip
  
  As a consequence, it holds also
  \be [D,L]=\ri g(\1-v\otimes v^\dagger).\label{787}\ee
  
  \medskip
  
Consider the following two-form $\omega$ defined on the manifold of the solutions $X,Y$
of the equation \eqref{765}
$$\omega=\tr(dY\wedge dX)=\tr\left(([k^\dagger dk,L]+dL)\wedge ([k^\dagger dk,D]+dD)\right)=$$
$$=\tr(dL\wedge dD) +\tr (k^\dagger dk\wedge d[L,D])+\jp\tr (k^\dagger dk \wedge [k^\dagger dk,[L,D]])=$$
\be =\tr(dL\wedge dD)=dp_j\wedge dq^j.\label{795}\ee
Note that the passage from the second to the third line of  Eq.\eqref{795} has used Eq.\eqref{787} in a substantial way (notably the fact that $v$ is the eigenvector of the unitary matrix $k$).

\medskip

So far we have shown that the space of solutions $X,Y$ of the equation \eqref{765} can be conveniently parametrized as $P_n\times U(n)_v$, where $P_n$ was defined in Eq.\eqref{740} and  $U(n)_v$ is the group of the unitary matrices which has $v$ as their eigenvector.
 Alternatively, we may parametrize the same space of solutions of Eq.\eqref{765} by first diagonalizing the Hermitian matrix $Y$, that is, by writing $X,Y$ as $X=\tilde k\tilde L \tilde k^\dagger$, $Y=-\tilde k\tilde D \tilde k^\dg$ where
 \be \tilde D={\rm diag} (\tilde q_1,...,\tilde q_n), \quad \tilde q_1 > \tilde q_2> ...>\tilde q_n,\label{803}\ee
  \be  \tilde L_{jk}= \tilde p^j\delta_{jk} -(1-\delta_{jk})\frac{\ri g}{\tilde q_j-\tilde q_k} \label{804b}\ee
 and $\tilde k\in U(n)_v$.
 
 \medskip
We  evaluate the form $\omega$ 
in the dual parametrization $X=k\tilde L k^\dagger$, $Y=k\tilde D k^\dg$, which gives 
\be \omega=\tr(dY\wedge dX)=d\tilde p^j\wedge d\tilde q_j.\label{825}\ee
The canonicity of the transformation
\eqref{813} is now the consequence of Eqs.\eqref{795}  and  \eqref{825}.

  \section{Lax integrability}
  
  In the context of infinite-dimensional phase spaces,   
  there appears no widely accepted definition of integrability in terms of  complete polarizations, and  the term "integrability"  often used in the infinite-dimensional literature means actually the so called {\it Lax integrability}. In fact,
the concept of the  Lax integrability exists also for the finite dimensional
integrable systems $(P,\A,\B)$ where it plays an important  technical role in the construction of the action-angle symplectomorphism 
  $\rho:P\to P$ such that $\B=\rho^*\A$. In the infinite number of dimensions, the Lax integrability gets promoted to a higher degree of importance and, de facto, it plays a prominent conceptual role there.
  
\subsection{The case of the finite number of degrees of freedom}
  
  Let $(P,H)$ be a finite-dimensional dynamical system, $V$ a  vector space and $\L(V)$ the space of linear operators on $V$. A Hamiltonian dynamical system $(P,H)$ is referred to as (strongly)
  Lax integrable if there exist  $\L(V)$-valued function
  $L$ on $P$ (called the Lax matrix) such that its spectral invariants generate a polarization $\B\subset \fp$ containing $H$. Recall in this respect that
  the spectral invariant of the operator $L\in\L(V)$ is any function $f:\L(V)\to \br$ such that
  $f(S^{-1}LS)=f(L)$ for  every invertible operator
  $S$ acting on $V$.
  
  \medskip
  
  In practice, it is useful to give sufficient conditions guaranteeing
  the Lax integrability. The most frequently used sufficient conditions amount to the existence
  of two more  matrix-valued functions $M:P\to\L(V)$ and $r:P\to\L(V\otimes V)$ such that
  \be \{L,H\}=[L,M],\label{843}\ee
  \be \{ L\otimes {\rm Id},\ld\}=[r,\lj]-[ r^p,\ld].\label{844}\ee
  In Eqs.\eqref{843} and \eqref{844},  Id is the identity operator on $V$, the notation $\{.,.\}$ stands for the Poisson brackets on $P$ and $[.,.]$ means the commutator
  of linear operators acting either on the space $V$ or on the space $V\otimes V$. If $r\in \L(V\otimes V)$ can be written as
  \be r=\sum_\alpha A_\alpha\otimes B_\alpha \ee
  for some family of linear operators $A_\alpha,B_\alpha\in \L(V)$, then the notation $r^p$ means  
  \be r^p=\sum_\alpha B_\alpha\otimes A_\alpha.\ee
 
  \medskip
  
  \noindent {\bf Remark 4.1}: {\small The fulfillment of the condition \eqref{843} guarantees   that the spectral invariants of the Lax operator $L$ commute with the Hamiltonian \cite{L} and it is sometimes referred to as the {\it weak} Lax integrability.
  On the other hand, the fulfillment of the condition \eqref{844} guarantees also that the spectral invariants commute among themselves \cite{S,M,BV}. In this case it is said that the dynamical system is {\it strongly} Lax integrable.}
  
 \medskip
  
  It is perhaps instructive to rewrite Eqs.\eqref{843} and \eqref{844} in some basis of the vector space $V$ in which the operators $L,M,r$ become matrices with appropriate subscripts and superscripts:
  \be \{L^i_{\ j},H\}=L^i_{\ k}M^k_{\ j}-M^i_{\ k}L^k_{\ j},\label{857}\ee
  \be \{L^i_{\ a}\delta^m_{\ b}, \delta^a_{\ j}L^b_{\ p}\}=r^{im}_{\ \ ab} L^a_{\ j} \delta^b_{\ p}-L^i_{\ a} \delta^m_{\ b}r^{ab}_{\ \ jp}-r^{mi}_{\ \ ba}\delta^a_{\ j}L^b_{\ p}+\delta^i_{\ a}L^m_{\ b}r^{ba}_{\ \ pj}.\label{868}\ee
  Of course, it is more convenient to rewrite Eq.\eqref{868} as
    \be \{L^i_{\ j}, L^m_{\ p}\}=r^{im}_{\ \ ap} L^a_{\ j} -r^{am}_{\ \ jp} L^i_{\ a}- r^{mi}_{\ \ bj} L^b_{\ p} +r^{bi}_{\ \ pj} L^m_{\ b}.\label{858}\ee

 \medskip
It can be easily verified   from Eq.\eqref{843} (or, equivalently, from Eq.\eqref{857}) that the spectral invariants of the Lax matrix $L$
are the constants of motion since they Poisson-commute with the Hamiltonian:
$$ \{\tr(L^s),H\}=\{L^{i_1}_{\ \ i_2}\dots L^{i_{s-1}}_{\ \ \  i_s} L^{i_s}_{\ \ i_1},H\}=$$$$ =\left(
L^{i_1}_{\ k}M^k_{\ i_2}-M^{i_1}_{\ k}L^k_{\ i_2}\right)\dots L^{i_{s-1}}_{\ \ \  i_s} L^{i_s}_{\ \ i_1}+\dots +L^{i_1}_{\ \ i_2}\dots L^{i_{s-1}}_{\ \ \  i_s}\left(
L^{i_s}_{\ k}M^k_{\ i_1}-M^{i_s}_{\ k}L^k_{\ i_1}\right)=
$$\be =\tr[L^s,M]=0.\ee
In a similar way, it follows from Eq.\eqref{844} (or, equivalently, from Eq.\eqref{858}) that the spectral invariants of the Lax matrix $L$
Poisson commute among themselves, that is
\be \{\tr(L^s),\tr(L^k)\}=0.\ee

\medskip

\noindent{\bf Remark 4.2} {\small
The existence of the matrices $L,M,r$  guarantees the  Poisson-commutativity of the eigenvalues of the matrix $L$
but it does not guarantee that those eigenvalues  be sufficiently numerous to generate
some polarization $\B$. Indeed, if the dimension of the vector space $V$ is small,  for example  just equal to $2$, and the dimension $2n$ of the phase space $P$
is  big, then the polarization, which  must be  locally generated by  $n$ functionally independent functions, cannot be  generated by just  two independent spectral invariants of the $2\times 2$-matrix $L$. In practice, the matrix $L$ must act on the vector space of sufficiently big dimension in order that its spectral invariants be able to generate a polarization. Let us illustrate this phenomenon on the example of
the Calogero-Moser system which is
  Lax integrable and the Lax matrix $L$ of which 
is given by the expression \eqref{769}
\be  L^j_{\ k}= p_k\delta^j_{\ k} -(1-\delta^j_{\ k})\frac{\ri g}{q^j-q^k}. \label{877}\ee
We have already established in Section 3.5 that the spectral invariants (i.e. the eigenvalues) of the matrix $L$ generate
the polarization $\B$ containing the Calogero-Moser Hamiltonian and
we now observe that the size $n\times n$ of the matrix $L$ is the minimal possible to accomplish this task. }

\medskip

For completeness, we give also the explicit expressions for the matrices $M$ and $r$ in the Calogero-Moser case 
 (cf. \cite{OP,AT, RuiBanff})
\be M^j_{\ k}= -(1-\delta^j_{\ k})\frac{\ri g}{(q^j-q^k)^2}+\delta^j_{\ k}\sum_{l\neq j}\frac{ig}{(q^j-q^l)^2},\ee
\be 
r^{im}_{\ \ kp}=\frac{1-\delta^{im}}{q^i-q^m}\delta^i_{\ p}\delta^m_{\ k}+\jp\delta^i_{\ k}\left( \frac{1-\delta^i_{\ p}}{q^p-q^i} \delta^{im}-\frac{1-\delta^{im}}{q^m-q^i}\delta^i_{\ p} \right).\label{882}\ee

\medskip

Technically, it is not always  practical to work with one Lax matrix of big size and it is sometimes more convenient to work with many 
  Lax matrices of small size  which together provide a sufficient number of Poisson-commuting constants of motion to generate a polarization of $P$.  The parameter $\lambda$  distinguishing the members of the family of those small Lax matrices is called the {\it spectral} parameter
and it often takes complex values.  Every single member of the
family of the Lax matrices is then denoted as $L(\lambda)$ and the dependence of this object on the phase space variable is   tacitly understood albeit  suppressed in the notations. 

\medskip

The concept of the Lax matrix with spectral parameter makes possible to generalize  the definition of the Lax integrable system given before.
The Hamiltonian dynamical system $(P,H)$ is thus referred to be
the   Lax integrable if there exists  a one-parameter family of $\L(V)$-valued functions
  $L(\xi)$ on $P$   such that the spectral invariants of all Lax matrices in the family  together generate   a polarization $\B\subset \fp$ containing $H$.

  \medskip

The eigenvalues of all matrices
  in the family $L(\xi)$  
  Poisson-commute with the Hamiltonian $H$ and Poisson-commute among themselves if there exist families of matrix-valued functions $M(\xi): P\to \L(V)$ and $r(\xi,\zeta):P\to\L(V\otimes V)$ such that  it holds
    \be \{L(\xi),H\}=[L(\xi),M(\xi)],\label{926}\ee
  \be \{L(\xi)\otimes {\rm Id},{\rm Id}\otimes L(\zeta)\}=[r(\xi,\zeta),L(\xi)\otimes {\rm Id}]-[r^p(\zeta,\xi),{\rm Id}\otimes L(\zeta)].\label{927}\ee
  
  \medskip
  
  We shall not  expand further  this text by reviewing finite dimensional systems
  which are Lax integrable by the Lax matrix with spectral parameter;
  the interested reader can find examples in the book \cite{BBT}. However, the Lax matrix with spectral parameter will be the main object
  of our interest in the next section devoted to the integrability of systems
  with infinite number of degrees of freedom. In that case, we do give
concrete examples of such Lax matrices, in particular those  relevant for the dynamics of $\E$-models. 

\subsection{Principal chiral model}

In what follows, we switch our attention  to the systems with infinite number of degrees of freedom. The criterion that the spectral invariants of the Lax matrix for all values of the spectral parameter should generate the polarization containing the Hamiltonian is now problematic to impose, because the very notion of the polarization is difficult to handle in the infinite dimensional case. For that reason one   requires instead that the Lax condition \eqref{926}   must encode the complete set of equations of motion of the model. Having in mind this circumstance, we concentrate in what follows on the issue of the Lax integrability
of the so called non-linear $\sigma$-models in $1+1$ spacetime dimensions 
which are field theories naturally associated to
the Riemannian-Kalb-Ramond manifolds. 

\medskip

Working in some coordinate patch $x^i$, we usually denote by $g_{ij}(x)$  the components of the  Riemannian tensor on $M$ and by $h_{ijk}(x)$ the components of the Kalb-Ramond field strength. Locally, $h_{ijk}(x)$ can be written in terms of the anti-symmetric Kalb-Ramond potential $b_{jk}(x)$ as 
\be  h_{ljk}(x)=  \d_lb_{jk}(x)+\d_jb_{kl}(x)+\d_kb_{lj}(x)\label{940}\ee
 and the $\sigma$-model action looks like (cf. Eq.\eqref{248})
 \be{  S[x(\tau,\sigma)]= \!\int  d\tau \oint \left(g_{ij}(x)+b_{ij}(x)\right)\d_+x^i\d_- x^j \equiv \!\int  d\tau \oint \ \! e_{ij}(x)\d_+x^i\d_- x^j .}\label{942}\ee

 The first known example of the Lax integrable $\sigma$-model was introduced in \cite{ZM} and it is know under the name of {\it principal chiral model}.  This theory can live on every quadratic Lie group $K$ but in the present paper $K$ will always stand for a  simple compact connected and simply connected Lie
 group.  The Riemannian structure on $K$ is proportional to  the bi-invariant metric equal to the (negative-definite) Killing-Cartan form $(.,.)_\K$ at the group origin and  there is no Kalb-Ramond field
 in this case.  Denoting the field configuration $k(\tau,\sigma)\in K$, the action of the principal chiral model is given by the expression  \be 
 S_0[k(\tau,\sigma)]=  -\int d\tau \oint ( k^{-1}\partial_+k,  k^{-1}\partial_-k)_\K.\label{946}\ee
 Occasionally, the action \eqref{946} is written in some suitable coordinates $x^i$ on the group manifold
 as follows
 \be S_0[x(\tau,\sigma)]= - \int d\tau \oint (E_i(x), E_j(x))_\K  \partial_+x^i   \partial_-x^j,\label{949}\ee
  where $E_i(x)dx^i$ is the $\K$-valued left-invariant Maurer-Cartan form on the group manifold $K$ expressed in the coordinates $x^i$.

  \medskip
  
  The phase space   $P=LK\times L\K$ of the principal chiral model is the direct product of the  loop group $LK$   with  the loop Lie algebra $L\K$. We denote the  points of $P$  as  pairs $(k(\sigma),\J(\sigma))\equiv (k,\J)$; the
  symplectic form is then given by the formula
  \be \omega_0= - \delta\oint  (\delta kk^{-1},\J)_\K   \label{958}
\ee
  and the positive definite Hamiltonian by 
\be H :=-\frac{1}{2}\oint (\d_\sigma kk^{-1},\d_\sigma kk^{-1})_\K-\frac{1}{2}\oint (\J,\J)_\K.\label{972}\ee
We associate to the data $(P,\omega_0,H)$
the first order action in the standard way:
\be S_0[k,\J]= \int d\tau\oint \left( -(\partial_\tau kk^{-1},\J)_\K+\jp  (\J,\J)_\K +  \jp (\partial_{\sigma}kk^{-1} ,\partial_{\sigma}kk^{-1} )_\K\right).\label{964}\ee
The current  $\J$ appears in this action  quadratically and can be easily eliminated to yield the second order $\sigma$-model action \eqref{946}.

\medskip

The Lax integrability of the principal chiral
model was established by Zakharov and Mikhailov in \cite{ZM}. The  Lax pair  $L(\xi),M(\xi)$ depending on a spectral parameter $\xi\in\bC$ is given by linear operators acting  on the loop Lie algebra $V\equiv L\K^\bc$ as follows
\be  L(\xi)=\partial_\sigma+\frac{1}{1-\xi^2}{\rm ad}_{(\xi\J-\d_\sigma kk^{-1})}, \qquad M(\xi)= \frac{1}{1-\xi^2}{\rm ad}_{(\xi\d_\sigma kk^{-1}-\J)}.\label{980}\ee
We must  verify that it holds  
\be \{L(\xi),H\}_0=[L(\xi),M(\xi)].\label{971x}\ee
To do that, it is necessary to determine the Poisson brackets $\{.,.\}$ generated by the symplectic form  \eqref{958}. They read
 \begin{subequations}  \label{999}
 \begin{align}
\label{999a}
 \{{\rm Ad}_{k(\sigma_1)}\otimes{\rm Id},{\rm Id}\otimes {\rm Ad}_{ k(\sigma_2)}\}_{0}&=0,
\\
\label{999b}
\{\Ad_{k(\sigma_1)},(\J(\sigma_2),x)_\K\}_{0}&=  -\ad_{x}\Ad_{k(\sigma_1)} \delta(\sigma_1-\sigma_2),\quad x\in\K
\\
\label{999d}
\{(\J(\sigma_1),x)_\K,(\J(\sigma_2),y)_\K\}_{0}&= -(\J(\sigma_1),[x,y])_\K\delta(\sigma_1-\sigma_2),\quad x,y\in\K.
\end{align}
\end{subequations}

\medskip

Let us now verify the relation \eqref{971x}. We first set
 \be \R:=\d_\sigma kk^{-1},\label{1027}\ee and then we infer from Eqs.\eqref{999} 
\be
\{(\J(\sigma_1),x)_\K,(\R(\sigma_2),y)_\K\}_{0}= -(\R(\sigma_1),[x,y])_\K\delta(\sigma_1-\sigma_2) -(x,y)_\K\delta'(\sigma_1-\sigma_2).
\ee

Then we find
\be\{L(\xi)(\sigma_1),(\J(\sigma_2),y)_\K\}_0
= \frac{[\xi\J(\sigma_1)-\R(\sigma_1),y] \delta(\sigma_1-\sigma_2)+ y \delta'(\sigma_1-\sigma_2)}{1-\xi^2},\label{1040}\ee
\be\{L(\xi)(\sigma_1),(\R(\sigma_2),y)_\K\}_0
= \frac{\xi[\R(\sigma_1),y] \delta(\sigma_1-\sigma_2)-\xi y \delta'(\sigma_1-\sigma_2)}{1-\xi^2},\label{1040b}\ee
 which gives the desired result
 \be \{L(\xi),H\}_0=-\frac{1}{1-\xi^2}\ad_{(\J-\xi\R)'+[\J,\R]}=[L(\xi),M(\xi)].\ee
Finally,  we find from Eqs.\eqref{980}, \eqref{1040} and \eqref{1040b} 
$$\{\tr(L(\xi)(\sigma_1)\ad_x),\tr(L(\zeta)(\sigma_2)\ad_y)\}=$$\be =\frac{(-\xi\zeta J(\sigma_1)+(\xi+\zeta)\R(\sigma_1),[x,y])_\K\delta(\sigma_1-\sigma_2)+(\xi+\zeta)(x,y)_\K\delta'(\sigma_1-\sigma_2)}{(1-\xi^2)(1-\zeta^2)},\label{1020a}\ee
The $r$-matrix satisfying the required relation
    \be \{L(\xi)\otimes {\rm Id},{\rm Id}\otimes L(\zeta)\}_0=[r(\xi,\zeta),L(\xi)\otimes {\rm Id}]-[r^p(\zeta,\xi),{\rm Id}\otimes L(\zeta)]\label{1021}\ee
    is therefore given by the expression
    \be r(\xi,\zeta)= \frac{\zeta^2}{1-\zeta^2}\frac{\C\delta(\sigma_1-\sigma_2)}{\xi-\zeta}\label{1024}.\ee
    
    \noindent {\bf Remark 4.3}: {\small The symbol $\C$ in Eq. \eqref{1024} stands for the Casimir element. The simplest
way to define $\C$ consists in picking a basis $T^a\in\K$ and define a Casimir matrix characterized by its matrix elements
\be C^{ab}:=(T^a,T^b)_\K\equiv \tr(\ad_{T^a}\ad_{T^b}).\label{1028e}\ee
Then in the defining relation of the Casimir element $\C$ appears the inverse Casimir matrix
\be \C:=C_{ab}\ \ad_{T^a}\otimes \ad_{T^b},\label{997}\ee
where  the Einstein summation convention applies. The Casimir element $\C$ defined in this way does not depend on the choice of the basis $T^a$.}

\bigskip

We note finally, that the first order Hamiltonian equations of motions of the principal chiral model can be written as
\be \partial_\tau \R=\{\R,H \}_0= \partial_\sigma \J +[\J,\R]; \quad \partial_\tau \J=\{\J, H \}_0= \partial_\sigma  \R.\ee
\section{Yang-Baxter deformations}
    \subsection{Yang-Baxter $\sigma$-model}
 The Yang-Baxter $\sigma$-model was introduced in Ref.\cite{K02} as the model exhibiting a particularly rich Poisson-Lie T-duality pattern and it was proven to be Lax integrable in Ref.\cite{K09}. In the present subsection, we give the definition of the model, we describe its Hamiltonian formalism and identify the Lax matrix $L(\lm)$ and the $r$-matrix $r(\lm,\zeta)$. 
 
 \medskip
 
 The Lagrangian of the Yang-Baxter $\sigma$-model is given by the formula
\be
 S_{  \beta}[k(\tau,\sigma)]=   -\frac{1}{2 \cos^4(\beta)} \int d\tau \oint ( k^{-1}\partial_+k, \frac{1}{1- \tan{(\beta)}R} k^{-1}\partial_-k)_\K.\label{1073}
\ee
Here $0\leq \beta <\frac{\pi}{2}$ denotes a deformation parameter and $R:\K^\bc\to\K^\bc$ is the so called Yang-Baxter operator
defined as
\be RE^{\al}=-{\rm sign}(\al){\rm i}E^{\al},\quad RH^j=0,\ee
where $E^\al,H^j$ is the standard  Chevalley basis   of   $\K^{\mathbb C}$. 

\medskip

The phase space   $P=LK\times L\K$ of the Yang-Baxter $\sigma$-model is the same as that of the  principal chiral one but   the symplectic form \eqref{958} is now  renormalized:
\be\omega_\beta=    -\frac{1}{ \cos^2{(\beta)}}\delta\oint (\delta kk^{-1},\J)_\K. \label{1081}  \ee
The Hamiltonian \eqref{972} turns out to be also deformed and it reads
\be H_\beta:=-\frac{1}{2}\oint (\R_\beta,\R_\beta)_\K-\frac{1}{2}\oint (\J,\J)_\K,\label{1085}\ee
where 
\be \R_{ \beta}:=\frac{1}{\cos^2{(\beta)}}{ \partial_{\sigma}kk^{-1}-}{ {\tan{(\beta)} }} { kR(k^{-1}\J k)k^{-1}}.\label{1087}\ee
If we associate to the data $(P,\omega_\beta,H_\beta)$
the first order action in the standard way
\be S_{\beta}[k,\J]= -{\frac{1}{ \cos^2(\beta)}} \int d\tau \oint (\partial_\tau kk^{-1},\J)_\K+\jp\int d\tau \oint  \left((\J,\J)_\K  +  (\R_{  \beta} ,\R_{  \beta})_\K\right)\ee
and integrate away the current $\J$, we recover the second order $\sigma$-model action \eqref{1073}. 

\medskip

The important fact is that the Hamiltonian equation of motions of the Yang-Baxter $\sigma$-model have the same form as those of the principal chiral model. Indeed, we find easily
\be \partial_\tau \R_\beta=\{\R_\beta,H_\beta \}_{\beta}=\d_\sigma\J+[\J,\R_\beta], \quad \partial_\tau \J=\{\J, H_\beta \}_{ \beta}= \d_\sigma \R_\beta,\label{1096}\ee
where $\{.,.\}_\beta$ is the Poisson bracket $\{.,.\}_0$ multiplied by the factor 
$\cos^2{(\beta)}$. The form of the equations \eqref{1096} suggests the Lax pair of the Yang-Baxter $\sigma$-model:
\be  L_{\beta}(\xi)=\partial_\sigma+\frac{1}{1-\xi^2}{\rm ad}_{(\xi\J-\R_\beta)}, \qquad M_\beta(\xi)= \frac{1}{1-\xi^2}{\rm ad}_{(\xi\R_\beta-\J)}.\label{1099}\ee
We must  verify that it holds indeed
\be \{L_\beta(\xi),H_\beta\}_\beta=[L_\beta(\xi),M_\beta(\xi)],\label{971y}\ee
which can be done straightforwardly by using the brackets of the currents $\J$ and $\R_\beta$:
{\small  \begin{subequations}  \label{1075}
 \begin{align}
\label{1075a}
\{(\J(\sigma_1),x)_\K,(\J(\sigma_2),y)_\K\}_{\beta}&= -\cos^2{(\beta)}\ (\J(\sigma_1),[x,y])_\K\delta(\sigma_1-\sigma_2), 
\\
\label{1075b}
\{(\R_\beta(\sigma_1),x)_\K,(\R_\beta(\sigma_2),y)_\K\}_{\beta}&= \phantom{-}\sin^2{(\beta)}\ (\J(\sigma_1),[x,y])_\K\delta(\sigma_1-\sigma_2),
\\
\label{1075c}
\{(\J(\sigma_1),x)_\K,(\R_\beta(\sigma_2),y)_\K\}_{\beta}&= -\cos^2{(\beta)}\ (\R_\beta(\sigma_1),[x,y])_\K\delta(\sigma_1-\sigma_2) -(x,y)_\K\delta'(\sigma_1-\sigma_2).
\end{align}
\end{subequations}}
 
Note that the brackets \eqref{1075} follow  from the brackets \eqref{999} as well as from the definition \eqref{1087}, due to the following important identities fulfilled by the Yang-Baxter operator $R$:
\be [Rx,Ry]=R[Rx,y]+R[x,Ry]+[x,y], \quad \forall x,y\in\K,\label{ybid}\ee
\be (Rx,y)_\K=-(x,Ry)_\K, \quad x,y\in\K.\label{antis}\ee

\medskip

We also find from Eqs.\eqref{1099} and \eqref{1075} 
{\small $$\{\tr(L(\xi)(\sigma_1)\ad_x),\tr(L(\zeta)(\sigma_2)\ad_y)\}=$$\be =\frac{((\sin^2{(\beta)}-\xi\zeta \cos^2{(\beta)}) J +(\xi+\zeta)\cos^2{(\beta)}\R ,[x,y])_\K\delta(\sigma_1-\sigma_2)+(\xi+\zeta)(x,y)_\K\delta'(\sigma_1-\sigma_2)}{(1-\xi^2)(1-\zeta^2)}.\label{1020b}\ee}
 The $r$-matrix $r_\beta(\xi,\zeta)$ satisfying the required relation
    \be \{L_\beta(\xi)\otimes {\rm Id},{\rm Id}\otimes L_\beta(\zeta)\}_\beta=[r_\beta(\xi,\zeta),L_\beta(\xi)\otimes {\rm Id}]-[r_\beta^p(\zeta,\xi),{\rm Id}\otimes L_\beta(\zeta)]\label{1128}\ee
    is then given by the expression
    \be r_\beta(\xi,\zeta)=  \phi_\beta^{-1}(\zeta)\frac{\C\delta(\sigma_1-\sigma_2)}{\xi-\zeta}\label{1024y},\ee
    where the function $\phi_\beta(\zeta)$, called the twist one in \cite{DMV}, is given by
    \be\phi_\beta(\zeta)=\frac{1-\zeta^2}{\cos^2{(\beta)}\zeta^2+\sin^2{(\beta)}}.\label{1131}\ee
    \subsection{$\E$-model formulation of the Yang-Baxter $\sigma$-model}
    It turns out that the action \eqref{1073} of the Yang-Baxter $\sigma$-model can be derived
    from an appropriate $\E$-model by the procedure detailed in Section 2.5. As the result, the action \eqref{1073} is the special
    case of the action \eqref{407}. Let us see how it works in more detail.
   
   \medskip
   
    For the underlying Drinfeld double $D$, we take the
    Lu-Weinstein one, that is $D$ is
    the special complex linear group ${ SL(n,\bc)}$ viewed as {\it real} group (i.e. it has the dimension $2(n^2-1)$ as the real manifold). 
The  non-degenerate symmetric ad-invariant bilinear form ${ (.,.)_{\D}}$ on the Lie algebra ${sl(n,\bc)}$ is defined by taking a suitable normalized  imaginary
 part of the trace
 \be {(X,Y)_{\D}= \frac{-1}{\sin{(\beta)}\cos^3{(\beta)}}\Im\tr(XY), \quad X,Y\in  sl(n,\bc)}.\label{bim}\ee
 The two half-dimensional isotropic  subgroups $K$ and $ \tilde K$ are
 respectively the special unitary group ${ SU(n)}$ and  the upper-triangular group ${ AN}$ with real positive numbers on the diagonal the product of which is equal to $1$.
 
 \medskip
 
 The one-parametric family of the $\E$-operators is given by
  \be \E_\beta X=-X+\frac{1+\ri\tan{(\beta)}}{2\ri\tan{(\beta)}}\left((1+\ri\tan{(\beta)})X+(1-\ri\tan{(\beta)})X^\dagger\right).\label{eb}\ee
  Recall that the second order $\sigma$-model action corresponding to the operator $\E_\beta$ reads
     \be S_\beta(k)=\jp\int d\tau \oint  
 \left(\left(E_\beta+\Pi(k)\right)^{-1}\partial_+kk^{-1}, \partial_-kk^{-1}\right)_\D.\label{2nd},\ee
where
  the linear operator $E_\beta:\tilde\K\to\K$ is defined by the operator $\E_\beta:\D\to\D$ in the way that   its graph $\{\tilde x+E_\beta\tilde x,\tilde x\in\tilde\K\}$ coincides with the image of the operator Id$+\E_\beta$. Moreover,
the $k$-dependent operator $\Pi(k):\tilde\K\to\K$   is given by  
\be \Pi(k)=-\J{\rm Ad}_k\tilde\J {\rm Ad}_{k^{-1}}\tilde\J,\label{PLS}\ee
where $\J,\tilde\J$ are projectors; $\J$ projects to $\K$ with the kernel $\tilde\K$ and  $\tilde\J$ projects to $\tilde\K$ with the kernel $\K$.

\medskip

To find out what is the  linear operator $E_\beta:\tilde\K\to\K$, it is convenient to parametrize the elements of the Lie algebra $\tilde\K$ by those of the Lie algebra $\K$ with the help of the Yang-Baxter operator $R$. Every element of $\tilde\K$ can be thus described in a unique way as $(R-\ri)y$, $y\in\K$ and we then find from Eq.\eqref{eb}
\be E_\beta(R-\ri)y=-Ry-\frac{y}{\tan{(\beta)}}.\label{eb'}\ee
Furthermore, from Eq. \eqref{PLS} we obtain
\be \Pi(k)(R-\ri)y=(R-\Ad_kR\Ad_{k^{-1}})y.\label{pk}\ee
Combining Eqs.\eqref{eb'} and \eqref{pk}, we find
\be (E_\beta+\Pi(k))^{-1}\d_+kk^{-1}=-(R-\ri)\frac{1}{\cot{(\beta)}+\Ad_kR\Ad_{k^{-1}}}\d_+kk^{-1}.\label{epi}\ee
Inserting \eqref{epi} into \eqref{2nd} and using \eqref{bim}, we recover indeed the action \eqref{1073} of the Yang-Baxter $\sigma$-model
\be
 S_{  \beta}(k)=   -\frac{1}{2 \cos^4(\beta)} \int d\tau \oint ( k^{-1}\partial_+k, \frac{1}{1- \tan{(\beta)}R} k^{-1}\partial_-k)_\K.\label{YBbis}
\ee
The fact that the  first-order Hamiltonian dynamics of the Yang-Baxter $\sigma$-model can be expressed in terms of the   $\E$-model \eqref{eb} explains the naturaleness of the current observables
$\J$, $\R_\beta$ employed in the previous section. Indeed, the crucial Poisson brackets \eqref{1075} can be obtained from the following general Poisson bracket derived from   the symplectic form \eqref{373}
\be \{(j(\si),X)_\D,(j(\si'),X')_\D\}=(j(\si),[X,X'])_\D+(X,X')_\D\d_\si\delta(\si-\si'),  \quad X,X'\in\D,\label{cur}\ee
where \be  j(\si)=\d_\si ll^{-1}.\ee
Working with the Lu-Wenstein double $K^\bc$ and the bilinear form \eqref{bim}, we recover from Eq.\eqref{cur} the Poisson brackets \eqref{1075} upon the identification
\be j(\si)=\cos{(\beta)}\bigl(\cos{(\beta)}\R_\beta(\si)+\ri\sin{(\beta)}\J(\si)\bigr).\label{ans}\ee
For example, setting $X=x$, $X'=\ri y$, $x,y\in\K$ in \eqref{cur} and using the ansatz \eqref{ans}, we recover the formula \eqref{1075c}.

\medskip

By the way, the formula \eqref{1087} can be also easily derived from the $\E$-model formalism. Indeed, decomposing
$l=k\tilde h$ like in Eq. \eqref{397}, we find
\be j=\d_\si ll^{-1}=\d_\si kk^{-1}+\Ad_k(\d_\si \tilde h\tilde h^{-1})=\d_\si kk^{-1}+\Ad_k((R-\ri)y), \quad y\in\K.\label{sost}\ee
Comparing Eqs.\eqref{ans} and \eqref{sost}, we recover the formula \eqref{1087}. Finally, we obtain
the Hamiltonian \eqref{1085} from the formulas \eqref{375}, \eqref{bim} and \eqref{ans}.
    \subsection{The   bi-YB-WZ $\sigma$-model and its special limits}

The Yang-Baxter $\sigma$-model introduced in the previous two sections is the prototypical representative of the family of the so-called Yang-Baxter integrable deformations of the principal chiral model. Since 2002, many multi-parametric Yang-Baxter deformations\footnote{ The integrable $\sigma$-models  constructed previously  on the target $SU(2)$ in Refs.\cite{C81,Mad93,Fa96,L12}  were later recognized to be of the Yang-Baxter type in \cite{K09,HRT,DHKM}.} were subsequently constructed e.g. in \cite{K14,DMV,Sf,KMY,OV,B,BW,GSS,DHKM,BL,K19,DHPT}. Here we review explicitely an example of the three-parametric  deformation living on the simple compact group target $K$  which  was introduced in Ref.\cite{DHKM,K19,K20} and bears the name of the bi-YB-WZ $\sigma$-model. The action of this model reads
     $$ S_{\rm bi-YB-WZ} 
(k)=\ka \int d\tau\oint \tr\biggl( k^{-1}\partial_+ k\frac{\al+e^{\rho_rR_{k}}e^{\rho_lR}}{\al-e^{\rho_rR_{k}}e^{\rho_lR}}k^{-1}\partial_- k\biggr) +$$\be +\ka\int \delta^{-1}\oint \tr(k^{-1}\delta k,[k^{-1}\partial_\sigma k,k^{-1}\delta k]).\label{biYBWZ}\ee 
Here
  the operator $R_k:\K\to\K$ is defined as
 \be R_k:=\Ad_k^{-1}R\Ad_k\label{Rm},\ee 
 furthermore     $\al\in]-1,1[$ and  $\rho_l,\rho_r\in]-\pi,\pi[$ are the deformation parameters and the real positive level $\ka$ is quantized as usual so that the WZ term exhibits the   $2\pi$ ambiguity.  Note that the case $\al=0$ corresponds to the standard WZNW model. 
  
  \medskip

Some other Yang-Baxter deformations previously constructed in the literature are appropriate special limits of the bi-YB-WZ $\sigma$-model \eqref{biYBWZ}. In particular, this is the case for the so called
YB-WZ $\sigma$-model 
  introduced in Ref.\cite{DMV15}. Several equivalent expressions were obtained for this YB-WZ  deformation in Refs.\cite{K17}, \cite{DDST18} and \cite{K19}. We reproduce here the   parametrization given  in \cite{K19}:
$$ S_{\rm YB-WZ} 
(k)=\ka \int d\tau\oint \tr\biggl( k^{-1}\partial_+ k\frac{\al+ e^{\rho_lR}}{\al- e^{\rho_lR}}k^{-1}\partial_- k\biggr) +$$\be +\ka\int \delta^{-1}\oint \tr(k^{-1}\delta k,[k^{-1}\partial_\sigma k,k^{-1}\delta k]).\label{YBWZ}\ee 
   Note that this is the special case of the action \eqref{biYBWZ} obtained by setting $\rho_r=0$.
   
   \medskip
   
 Furthermore, setting \be \rho_r=2\ka b_r,\quad \rho_l=2\ka b_l,\quad \al=e^{-2\ka a},\label{par}\ee
and taking limit $\ka\to 0$, we recover from  the action \eqref{biYBWZ} the bi-Yang-Baxter integrable deformation of the principal chiral model introduced in \cite{K02,K14}:
 \be S_{\rm bi-YB}(k)=-\int d\tau\oint \tr\biggl( k^{-1}\partial_+ k \left(a+b_rR_k +b_lR\right)^{-1}k^{-1}\partial_- k\biggr).\label{biYB}\ee

 Taking moreover $b_r=0$, we recover the Yang-Baxter $\sigma$-model \eqref{YBbis} 
 \be S_{\rm YB}(k)=-\int d\tau\oint \tr\biggl( k^{-1}\partial_+ k \left(a  +b_lR\right)^{-1}k^{-1}\partial_- k\biggr),\label{YB}\ee
 with $a=2\cos^4{(\beta)}$, $b_L=-2\sin{(\beta)}\cos^3{(\beta)}$.
      \subsection{$\E$-model formulation of the bi-YB-WZ $\sigma$-model}
  Recall that the $\E$-model is the first order
  dynamical system living on the loop group of the Drinfeld double where the symplectic form and the Hamiltonian read respectively
   \be   \omega =-\jp\oint   \bigl(l^{-1}\delta l\stackrel{\wedge}{,}(l^{-1}\delta l)'\bigr)_{\D}\label{esf},\ee
 \be H_\E  =\jp\oint \bigl(l'l^{-1},\E \ \! l'l^{-1}\bigr)_{\D}\equiv \jp\oint \bigl(j,\E \ \! j\bigr)_{\D} .\label{eha}\ee
 The Poisson brackets of the currents $j=l'l^{-1}$ derived from the symplectic form
 \eqref{esf} then read
 \be \{(j(\si),X)_\D,(j(\si'),X')_\D\}=(j(\si),[X,X'])_\D+(X,X')_\D\d_\si\delta(\si-\si'),  \quad X,X'\in\D.\label{curbis}\ee
 Whenever there is given a maximally isotropic
 Lie subalgebra $\tilde\K\subset\D$, we can write down the second order $\sigma$-model action 
$$S_\E(l)=\frac{1}{4}\int \delta^{-1}\oint \biggl(\delta ll^{-1},[\partial_\sigma ll^{-1},\delta ll^{-1}]\biggr)_\D + $$\be +\frac{1}{4} \int d\tau\oint \biggl( \partial_+ ll^{-1},Q^-_l \partial_- ll^{-1}\biggr)_\D-\frac{1}{4} \int d\tau\oint \biggl(Q^+_l\partial_+ ll^{-1}, \partial_- ll^{-1}\biggr)_\D.\label{dokbis} \ee
 Here $l(\tau,\sigma)\in D$ is a field configuration and $Q^\pm_l:\D\to\D$ are the projectors  fully characterized by their respective images  and kernels
\be \label{projectorsbis}  \mathrm{Im}(Q_l^\pm)=Ad_l(\tilde\K),  \quad
  \mathrm{Ker}(Q_l^\pm) =(1\pm\E)\D. \ee
  The upshot is that the first order Hamiltonian dynamics of the $\sigma$-model \eqref{dokbis} is given by the data \eqref{esf}, \eqref{eha} and \eqref{curbis}.
  
  \medskip
  
It was shown in Ref.\cite{K20} that the first order Hamiltonian dynamics of the bi-YB-WZ $\sigma$-model \eqref{biYBWZ} can be formulated in terms of an appropriate $\E$-model. Indeed, the  underlying Drinfeld double is  
 $D=K^\bc$ as in the one-parametric Yang-Baxter case, however, the bilinear form $(.,.)_\D$ now reads
 \be \left(X,X'\right)_\D:=\frac{4\ka}{\sin{(\rho_l)}}  \Im\tr\left(e^{\ri \rho_l}XX'\right),\quad X,X'\in \K^\bc.\label{bimbis}\ee
Here the symbol  $\Im$ means  the imaginary  part of a complex number and  $\ka,\rho_l$ are the parameters appearing in the
action \eqref{biYBWZ}.  

\medskip

The operator $\E$ is given by the formula
\be \E_{\al,\rho_l,\rho_r}X= -X+\frac{1}{\al-\al^{-1}}(\al -e^{-\ri\rho_l}e^{-\rho_rR})\left(X-X^\dagger-\frac{\cos{(\rho_l)}-\al^{-1}e^{\rho_rR}}{\sin{(\rho_l)}}\ri(X+X^\dagger)\right),\label{1333}\ee
where again the parameters $\al,\rho_l$ and $\rho_r$ are those appearing in the action \eqref{biYBWZ}.

\medskip

The maximally isotropic Lie subalgebra  $\tilde\K\subset K^\bc$ defined as
\be \tilde\K=\left\{ \frac{e^{-\ri \rho_l}-e^{-\rho_l R}}{\sin{\rho_l}}y, \ y\in\K\right\}.\label{mis}  \ee
The fact that the subspace of $\D$ defined by Eq.\eqref{mis} is indeed the Lie subalgebra of $\D$ is the consequence of the properties of the Yang-Baxter operator, namely of the identity \eqref{ybid} rewritten as
 \be \left[\frac{e^{-\ri \rho_l}-e^{-\rho_l R}}{\sin{\rho_l}}x,
\frac{e^{-\ri \rho_l}-e^{-\rho_l R}}{\sin{\rho_l}}y\right]=
\frac{e^{-\ri \rho_l}-e^{-\rho_l R}}{\sin{\rho_l}}[x,y]_{R,\rho_l},\quad x,y\in \K.\label{Bbracket}\ee
Here $[.,.]_{R,\rho_L}$ is a new Lie bracket on the vector space $\K$  which is defined with the help of the standard Lie bracket $[.,.]$ and of the Yang-Baxter operator as
\be [x,y]_{R,\rho_l}:=\left[\frac{\cos{ \rho_l}-e^{-\rho_l R}}{\sin{\rho_l}}x,y\right]+\left[x,\frac{\cos{ \rho_l}-e^{-\rho_l R}}{\sin{\rho_l}}y\right].\label{Rrhobracket}\ee

\medskip

As it was explained in Ref.\cite{K20}, the subgroup $\tilde K\subset D$ corresponding to the Lie subalgebra $\tilde\K$ turns out to be the semi-direct product of a suitable real form of the complex Cartan torus $\mathbb T^\bc$ with the nilpotent subgroup $N\subset K^\bc$ generated  by the positive step operators
$E^\al$.   The space of cosets then $D/\tilde K$ turns out to be just the group $K$, therefore the gauge fixing
$l=k$ transforms the model \eqref{dokbis} into some $\sigma$-model living on the group $K$. To see which one, we must evaluate the expressions $Q_k^\pm\d_\pm kk^{-1}$. To do that we first find from Eq.\eqref{1333}
\be (1\pm\E)\D   =\left\{(\al^{\pm 1}-e^{-\ri\rho_l}e^{-\rho_rR})x, \  x\in\K\right\}\label{epm}\ee
and then we write down the identity 
$$\d_\pm kk^{-1}=\left(e^{-\ri \rho_l}-e^{-\rho_l R_{k^{-1}}}\right)\left(\alpha^{\pm 1} e^{\rho_rR}-e^{-\rho_lR_{k^{-1}}}\right)^{-1}\d_\pm kk^{-1}+$$\be +(\al^{\pm 1}-e^{-\ri\rho_l}e^{-\rho_rR})(\al^{\pm 1}-e^{-\rho_lR_{k^{-1}}}e^{-\rho_rR})^{-1}\d_\pm kk^{-1}.\label{ide}\ee
Taking into account the definition \eqref{projectorsbis} of the projectors $Q^\pm_k$ as well as Eqs.\eqref{mis},\eqref{epm} and \eqref{ide}, we infer
 \be Q_k^\pm\d_\pm kk^{-1}=\left(e^{-\ri \rho_l}-e^{-\rho_l R_{k^{-1}}}\right)\left(\alpha^{\pm 1} e^{\rho_rR}-e^{-\rho_lR_{k^{-1}}}\right)^{-1}\d_\pm kk^{-1}.\label{cvy}\ee
 Inserting the expressions \eqref{cvy} into the action \eqref{dokbis} written for $l=k$ we find   precisely the 
 bi-YB-WZ action
     $$ S_{\rm bi-YB-WZ} 
(k)=\ka \int d\tau\oint \tr\biggl( k^{-1}\partial_+ k\frac{\al+e^{\rho_rR_{k}}e^{\rho_lR}}{\al-e^{\rho_rR_{k}}e^{\rho_lR}}k^{-1}\partial_- k\biggr) +$$\be +\ka\int \delta^{-1}\oint \tr(k^{-1}\delta k,[k^{-1}\partial_\sigma k,k^{-1}\delta k]).\label{biYBWZbis}\ee

  \subsection{Lax integrability of the bi-YB-WZ $\sigma$-model}
 
The Lax 
   pair operators $L(\xi)$ and $M(\xi)$ of the bi-YB-WZ $\sigma$-model  were identified in Ref.\cite{K20}. In this paper, we rewrite them in an equivalent but simpler   way as follows
    \be L(\xi)  =\d_\si - \ad_{O(\xi)j},\label{Laxfh}\ee
    \be M(\xi)=- \ad_{O(\xi)\E j}.\label{M}\ee
Here the involution $\E$ is given by Eq.\eqref{1333},  the $\br$-linear operator $O(\xi):\D\to\D$ is defined as
\be O(\xi)j=(1+h(\xi)e^{-\rho_rR})\frac{e^{\ri\rho_l}j+e^{-\ri\rho_l}j^*}{2\ri\sin{(\rho_l)}}+g(\xi)\frac{j+j^*}{2\ri\sin{(\rho_l)}}\label{oj}\ee
and the meromorphic functions $h(\xi)$, $g(\xi)$ must verify the relation 
      \be  h^2(\xi)+  g^2(\xi)+ (\al+\al^{-1})h(\xi)g(\xi)+2 \cos{(\rho_l)}g(\xi)+2\cos{(\rho_r)}h(\xi)+1=0.\label{pqid}\ee 
      We can now straighforwardly verify the first condition \eqref{926} of the Lax integrability, that is the validity of the
  relation
       \be  \{L(\xi),H_\E\}= \{L(\xi),\jp\oint \bigl(j,\E \ \! j\bigr)_{\D} \}=[L(\xi),M(\xi)].\label{Laxcondition}\ee
  Using the current Poisson brackets \eqref{curbis} and the definition of the Hamiltonian \eqref{eha}, we first find 
  \be \{j,H_\E\}=\E\d_\si j+[\E j,j],\label{sostbis}\ee
  which means that for verifying the condition \eqref{Laxcondition} we must 
  just prove the identity\footnote{The identity \eqref{sid} appeared already in Ref.\cite{LV20} as the reformulation of the identity introduced in Ref.\cite{S17}.} 
  \be O(\xi)[\E j,j]=[O(\xi)\E j, O(\xi)j].\label{sid}\ee
The  validity of the identity \eqref{sid} is then the straightforward consequence of the relation \eqref{pqid} as well as of  the following identity proved in Ref.\cite{K20} \be [e^{\rho_rR}x,e^{\rho_r R}y]= e^{\rho_r R}\left( [e^{\rho_r R}x,y]+[x,e^{\rho_r R}y]-2\cos{(\rho_r)}[x,y]\right) +[x,y],\quad x,y\in \K.\label{id}\ee
   By the way, for the meromorphic functions $g(\xi)$ and $h(\xi)$ verifying 
the condition \eqref{pqid}, we can take e.g.
  \be g(\xi)=\frac{4\cos{(\rho_l)}-2\cos{(\rho_r)}(\al+\al^{-1})}{(\al-\al^{-1})^2}+\frac{ 2A\xi}{1-\xi^2}+\frac{A(\al+\al^{-1})}{(\al-\al^{-1})} \frac{1+\xi^2}{1-\xi^2},
   \ee
       \be h(\xi)=\frac{4\cos{(\rho_r)}-2\cos{(\rho_l)}(\al+\al^{-1})}{(\al-\al^{-1})^2} -\frac{2A}{(\al-\al^{-1})} \frac{1+\xi^2}{1-\xi^2},
   \ee   
   where
   \be A=\sqrt{1+4\times\frac{\cos^2{(\rho_l)}+\cos^2{(\rho_r)}-(\al+\al^{-1})\cos{(\rho_l)}\cos{(\rho_r)}}{(\al-\al^{-1})^2}}.\ee
       Recall that the second condition \eqref{927} of the Lax integrability reads
        \be \{L(\xi)\otimes {\rm Id},{\rm Id}\otimes L(\zeta)\}=[r(\xi,\zeta),L(\xi)\otimes {\rm Id}]-[r^p(\zeta,\xi),{\rm Id}\otimes L(\zeta)].\label{llrbis}\ee
        Fixing a basis $T^a\in\K$, we make the following ansatz for the $r$-matrix
        \be r(\xi,\eta)=C_{ab}\ad_{\hat r(\xi,\eta)T^a}\otimes\ad_{T^b}\delta(\si_1-\si_2),\ee
       so that the operator $\hat r(\xi,\eta):\K\to\K$ is the quantity  that we wish to find.
       
 Contracting the condition \eqref{llrbis} with an element $\ad_x\otimes \ad_y$, $x,y\in\K$,  considering $\xi,\zeta\in\br$ and using Eq.\eqref{Laxfh}, we obtain
 $$\{(O(\xi)j(\si_1),x)_\K,(O(\zeta)j(\si_2),y)_\K\}=$$$$ = -\Bigl((O(\xi)j(\si_1),[x,\hat r(\xi,\zeta)y])_\K+(O(\zeta)j(\si_2),[\hat r(\zeta,\xi)x,y])_\K\Bigr)\delta(\si_1-\si_2)$$\be -\Bigl((x,\hat r(\xi,\zeta)y)_\K+(\hat r(\zeta,\xi)x,y)_\K\Bigr)\delta'(\si_1-\si_2).\label{por}\ee
 Note that for $\xi\in\br$, the image of the operator $O(\xi)$ is just the Lie algebra $\K$, moreover, it exists the operator $O^\dagger(\xi):\K\to\D$, which is "adjoint" to $O(\xi):\D\to\K$ in the
 sense of the equality
 \be (O(\xi)j,x)_\K=(j,O^\dagger(\xi)x)_\D.\label{ado}\ee
 Indeed, $O^\dagger(\xi)$ is given by the formula
\be O^\dagger(\xi)x=\frac{(\I+h(\xi)e^{\rho_rR})x+e^{-\ri \rho_l}g(\xi)x}{4\ka}.\label{dag}\ee
 Now because of the relations \eqref{ado} and \eqref{curbis}, we have also
  $$\{(O(\xi)j(\si_1),x)_\K,(O(\zeta)j(\si_2),y)_\K\}=$$\be =(j,[O^\dagger(\xi)x,O^\dagger(\zeta)y])_\D  +(O^\dagger(\xi)x,O^\dagger(\zeta)y)_\D\d_\si\delta(\si-\si'),  \quad x,y\in\K.\label{curtris}\ee
  Finally, comparing Eq.\eqref{por} with Eq.\eqref{curtris}, we infer that the operator $\hat r$ must fulfil the following conditions
  \be [O^\dagger(\xi)x,O^\dagger(\zeta)y]=-O^\dagger(\xi)[x,\hat r(\xi,\zeta)y]-O^\dagger(\zeta)[\hat r(\zeta,\xi)x,y],\label{1c}\ee
  \be (O^\dagger(\xi)x,O^\dagger(\zeta)y)_\D= -(x,\hat r(\xi,\zeta)y)_\K-(\hat r(\zeta,\xi)x,y)_\K.\label{2c}\ee
 Using the identities \eqref{pqid} and \eqref{id},  it is then straightforward to work out that the conditions \eqref{1c} and \eqref{2c} are fulfilled
 by the following operator $\hat r(\xi,\eta)$
         \be \hat r(\xi,\zeta)=\frac{h(\zeta)}{4\ka}\Bigl(w(\xi,\zeta){\rm Id}- e^{\rho_r R}\Bigr),\label{biYBWZrmatrix}\ee  
         where 
        \be w(\xi,\zeta)=  \frac{h(\xi)g(\zeta)+h(\zeta)g(\xi)+ (\al+\al^{-1}) g(\xi)g(\zeta)+2\cos{(\rho_r)}g(\xi)}{g(\xi)-g(\zeta)}.\label{v}\ee
 The strong Lax integrability of the bi-YB-WZ model is thus established. 
 
 \subsection{Sufficient conditions of the strong Lax integrability}
 Sufficient conditions for the weak Lax integrability of non-linear $\sigma$-models were formulated in Refs.\cite{M02},\cite{M08}
 and they turned out to be useful for finding new examples of the integrable $\sigma$-models in Ref.\cite{M20}. As far as  the weak Lax integrability of the $\E$-models is concerned, the sufficient conditions were formulated in Ref.\cite{S17} and they turned out to be useful for constructing new examples of integrable $\E$-models in Ref.\cite{LV20}. Actually, the results of Section 5.5 show that the weak integrability condition
 of Refs.\cite{S17,LV20} can be supplemented by further conditions imposed on the family of operators $O(\xi)$ which guarantee also the strong Lax integrability of the theory. Indeed, if for a given $\E$-model  we find the families of linear operators $O(\xi)$, $O^\dagger(\xi)$ and
 $\hat r(\xi,\zeta)$ verifying the conditions 
   \be O(\xi)[\E j,j]=[O(\xi)\E j, O(\xi)j],\label{sidbis}\ee
    \be [O^\dagger(\xi)x,O^\dagger(\zeta)y]=-O^\dagger(\xi)[x,\hat r(\xi,\zeta)y]-O^\dagger(\zeta)[\hat r(\zeta,\xi)x,y],\label{1cbis}\ee
  \be (O^\dagger(\xi)x,O^\dagger(\zeta)y)_\D= -(x,\hat r(\xi,\zeta)y)_\K-(\hat r(\zeta,\xi)x,y)_\K,\label{2cbis}\ee
 then the $\E$-model is strongly Lax integrable. The results of Section 5.5 can be then also interpreted in the way, that the general sufficient conditions \eqref{sidbis},\eqref{1cbis} and \eqref{2cbis} of the strong Lax integrability have the particular nontrivial solution given by the operators \eqref{1333}, \eqref{oj}, \eqref{dag} and 
 \eqref{biYBWZrmatrix}.
 
 \vskip3pc
 
 \noindent {\bf Acknowledgement}: I am indebted to Simon Ruijsenaars for reading the manuscript and suggesting several important improvements in particular in Section 3.

  \end{document}